\documentclass[bibyear]{aa} 
\usepackage{amsmath}
\usepackage{amssymb}
\usepackage{graphics}
\usepackage{graphicx}
\usepackage{natbib}
\usepackage{gensymb}
\usepackage{pdflscape}
\usepackage{capt-of}
\usepackage{float}
\usepackage{txfonts}
\usepackage{xcolor}
\usepackage[caption = false]{subfig}

\authorrunning{Sharma et al.}

\begin{document}

   \title{Wave amplitude modulation in fan loops as observed by AIA/SDO}


   \author{Aishawnnya Sharma
          \inst{1,2},
         Durgesh Tripathi\inst{1},
         R. Erd\'elyi\inst{3, 4},
         G. R. Gupta\inst{5} 
         \and
          Gazi A. Ahmed\inst{2}
          }

   \institute{Inter-University Centre for Astronomy and Astrophysics, Post Bag-4, Ganeshkhind, Pune 411007, India\\
   \and
   Department of Physics, Tezpur University, Tezpur 784028, India\\
   \and
        Solar Physics and Space Plasma Research Centre (SP$^2$RC), School of Mathematics and Statistics, University of Sheffield, Hounsfield Road, Hicks Building, Sheffield, S3 7RH, UK  \\
        \and
        Department of Astronomy, E\"otv\"os Lor\'and University, Budapest, P\'azm\'any P. s\'et\'any 1/A, H-1117, Hungary\\
        \and
        Physical Research Laboratory, Udaipur Solar Observatory, Devali, Badi Road, Udaipur 313001, India\\
             \email{aish@iucaa.in, sharmaaishawnnya@gmail.com}      }


  \abstract
   {}
   {To perform a detailed analysis to understand the evolution and dynamics of propagating intensity disturbances observed in a fan loop system.}
  {We perform multiwavelength time-distance analysis of a fan loop system anchored in an isolated sunspot region (\textit{AR 12553}). The active region was observed by the Atmospheric Imaging Assembly on-board the Solar Dynamics Observatory. We measure the phase speeds of the propagating intensity disturbances by employing cross-correlation analysis, as well as by obtaining the slopes in xt-plots. We obtain original as well as de-trended light curves at different heights of the time-distance maps and characterize them by performing Fouri\'er and Wavelet analysis, respectively.}
   {The time-distance maps reveal clear propagation of intensity oscillations in all the coronal EUV channels except AIA 94 and 335~{\AA}. We determine the nature of the intensity disturbances as slow magneto-acoustic waves by measuring their phase speeds. The time-distance maps, as well as the de-trended light curves, show an increase and decrease in the amplitude of propagating 3-min oscillations over time. The amplitude variations appear most prominent in AIA 171~{\AA}, though other EUV channels also show such signatures. Fouri\'er power spectrum yield the presence of significant powers with several nearby frequencies between 2--3 minutes (5--8 mHz), along with many other smaller peaks between 2--4 minutes. Wavelet analysis shows an increase and decrease of oscillating power around 3-min simultaneous to the amplitude variations. We obtain the modulation period to be in the range of 20--30 minutes.}
   {Our results provide the viability of occurrence of phenomenon like `Beat' among the nearby frequencies giving rise to the observed amplitude modulation. However, we can not at this stage rule out the possibility that the modulation may be driven by variability of an underlying unknown source.}

\maketitle
%
\section{Introduction}\label{intro}
Identification and characterization of magnetohydrodynamic (MHD) waves and oscillations are major thrust areas of research in the field of solar plasma-astrophysics. MHD waves are considered to be an important candidate for heating of the solar atmosphere \citep{2004A&G....45d..34E, 2005SSRv..120...67O, 2009SSRv..149..229T, 2019ApJ...871..155R, Liu2019}. They may have significant roles in triggering of jets \citep{2004Natur.430..536D, 2015MNRAS.446.3741C} as well as in the acceleration of fast  solar wind \citep{2009LRSP....6....3C}.

One of the wave modes observed in the solar corona is slow mode magneto-acoustic waves. The propagation of this wave mode is largely confined to the magnetic field direction. They are compressional in nature, and propagate with speed that is less than the local sound speed in the solar corona. Such waves are often observed as fluctuations in intensity as well as in Doppler velocity \citep{2009A&A...503L..25W, 2012A&A...546A..93G}. These waves were first reported as propagating intensity disturbances (PIDs) by \citet{1997ApJ...491L.111O} in coronal plumes in off-limb observations recorded by the UltraViolet Coronagraph Spectrometer \citep[UVCS,][]{1995SoPh..162..313K} on-board  the \textit{Solar and Heliospheric Observatory (SOHO)}. Similar intensity oscillations in coronal fan loops were first detected by \citet{1999SoPh..186..207B} using the Extreme-ultraviolet Imaging Telescope \citep[EIT,][]{1995SoPh..162..291D} on board SOHO in 195~{\AA} channel, and later confirmed by \citet{2000A&A...355L..23D} in the observations recorded by Transition Region and Coronal Explorer \citep[TRACE,][]{1999SoPh..187..229H} in the 171~{\AA} channel. Fan loops are magnetic plasma structures in the corona with temperature between 0.6{--}1 MK \citep{2010LRSP....7....5R, 2017ApJ...835..244G}. They are found at the edges of active regions, and are mostly rooted in the umbra and umbra-penumbra boundaries of sunspots. They also remain anchored to the penumbra and non-sunspot (such as, plage) regions. 
From a statistical study, \citet{2012SoPh..279..427K} concluded that PIDs of sunspot origin are temperature dependent and have subsonic propagation speed.

Since the discovery of PIDs in coronal fan loops, a number of theoretical and observational studies have been performed to unravel their characteristics, propagation and dissipation and also their counterparts in the lower solar atmosphere \citep[see, e.g.,][]{2000A&A...362.1151N, 2002SoPh..209...89D, 2002SoPh..209...61D, 2006RSPTA.364..447R, 2009SSRv..149...65D, 2000SoPh..192..373B, 2006RSPTA.364..313B, 2015LRSP...12....6K}. Using wavelet analysis, \citet{2002A&A...387L..13D} found that the periods of the PID are of the order of $282\pm93$ s. However, careful observation of their footpoint locations showed that the coronal fan loops rooted in sunspot regions have a dominant period of $172\pm32$ s, and the loops rooted in non-sunspot regions, such as plages are dominant in $321\pm74$ s period oscillations. This distinction was explained in terms of the underlying driver at the photospheric level, exciting the loop footpoints.

In addition to the regular 3-min and 5-min oscillation, few studies have reported longer-period oscillations in the fan loops. \citet{2009A&A...503L..25W} has reported longer-period propagating waves at around 12 and 25 minutes along a fan-like coronal structure in the intensity \citep[see also][]{2009ApJ...697.1674M} and Doppler shift of the Fe XII~195 ~{\AA} line using EUV Imaging Spectrometer \citep[EIS,][]{2007SoPh..243...19C} on-board \textit{HINODE} \citep{kosugi2007}. \citet{2011A&A...526A..58S} showed the presence of waves in multiples of 8 minutes in coronal fan loops such as, in 16-min, 24-min, 40-min etc. It was speculated that the presence of the 8-min period could be attributed to beat phenomenon, which may occur due to the superposition of ubiquitous 3-min chromospheric and 5-min photospheric oscillations that leak into the corona through the inclined guide field  that is provided by fan loops. However, they did not emphasise the excitation mechanisms of other longer-period oscillations that they detected in their observations.

In this paper, we perform a detailed analysis of intensity perturbations propagating along a fan loop system anchored within a sunspot. We track the evolution of these intensity oscillations by using the high spatially and temporally resolved data recorded by the Atmospheric Imaging Assembly \citep[AIA,][]{2012SoPh..275...17L} on-board the \textit{Solar Dynamic Observatory} \citep[SDO,][]{2012SoPh..275....3P}. Our motivation is to understand how intensity oscillations with a range of periodicities interact, and affect the propagation characteristics of most dominant 3-min oscillations along fan loops. The rest of the paper is structured as follows. In Section \ref{obs}, we describe the observations obtained for this study and briefly address the data reduction and preparation. We present the details of the different analyses acquired in this study, such as time-distance maps, Fouri\'er analysis, wavelet analysis and respective results in Section \ref{analysis}. Finally, we summarise our findings and conclude in Section \ref{summary}.
\section{Observations}\label{obs}
\begin{figure*}[!hbtp]
\centering
\includegraphics[trim=1.0cm 4.0cm 0.0cm 4.0cm, width=0.65\textwidth]{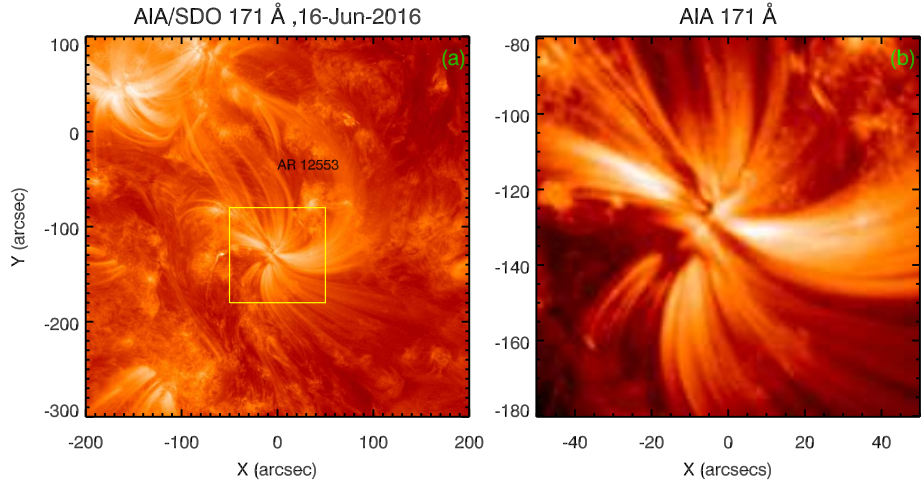}
\caption{(a) AIA 171~{\AA} image showing the active region. The yellow box encloses the fan loop system, the region of interest, that is studied in this paper. (b) Zoomed-up image corresponding to the yellow box shown in panel (a).}
\label{fig:ss171}
\end{figure*}
The main aim of this work is to perform a multi-wavelength study of the dynamics of a fan loop system. For this purpose we have utilised the observations recorded by AIA using its UV channel, AIA 1600~{\AA} and all the EUV channels, AIA 304~{\AA}, 171~{\AA}, 131~{\AA}, 193~{\AA}, 211~{\AA}, 335~{\AA}, and 94~{\AA}. These channels are dominated with different spectral lines formed in a range of temperatures. For more information on the temperature sensitivity of different channels see \citet{O'Dwyer2010}, \citet{DelZanna2011}, and \citet{2012SoPh..275...41B}. AIA provides full disk images with a cadence of 12~s in EUV and 24~s in UV. In addition, for contextual purpose, we use continuum data obtained using the Helioseismic and Magnetic Imager \citep[HMI,][]{2012SoPh..275..327S, 2012SoPh..275..229S}, also on-board SDO. The AIA and HMI data are processed using standard processing software provided in the solar software \citep[SSW;][]{1998SoPh..182..497F} distribution. The processed level~1.5 AIA and HMI data provides a pixel resolution of  0.6{\arcsec}. 
\begin{figure*}[!hbtp]
\centering
\includegraphics[width=0.8\textwidth]{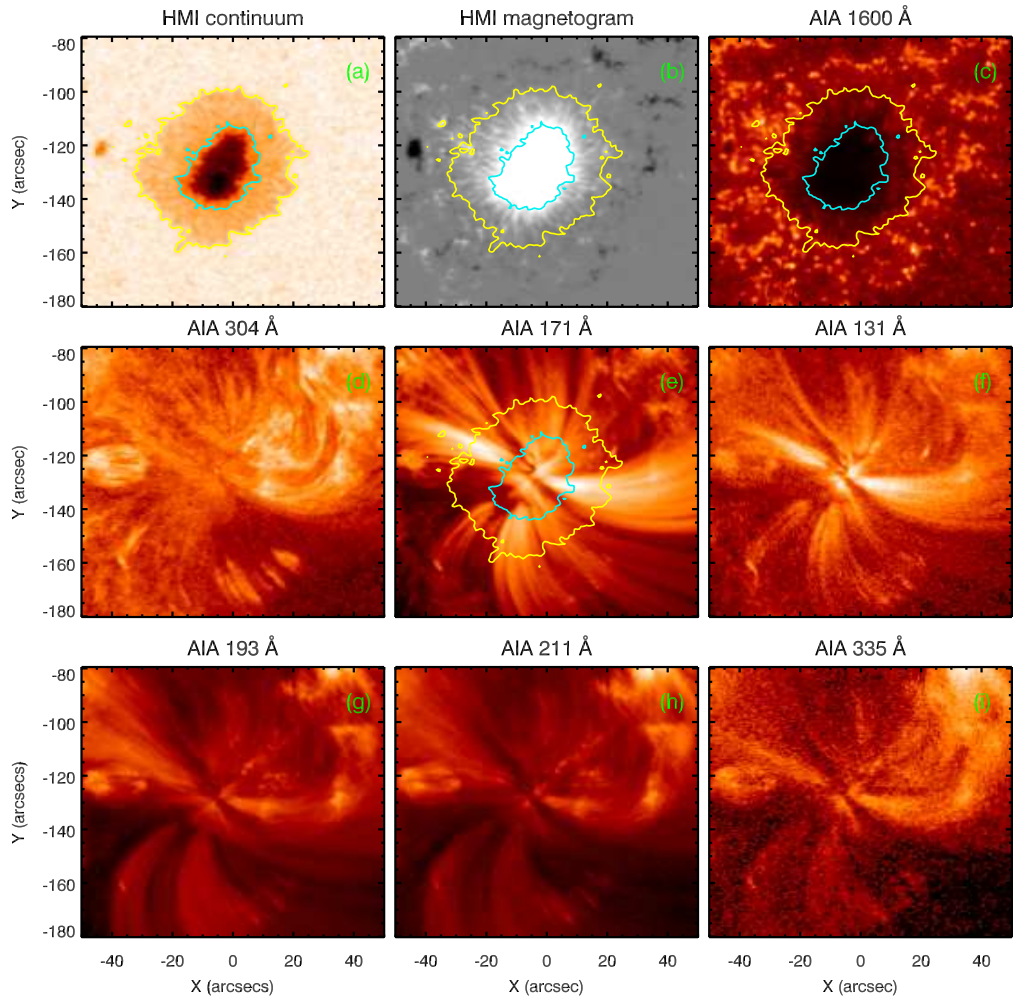}
\caption{Regions of interest observed with HMI and AIA in different wavelengths as labelled. The overlaid cyan and yellow contours in panels (a), (b), (c) and (e) are the umbra-penumbra and penumbra-outer sunspot boundaries, respectively obtained from HMI continuum.}
\label{fig:sunspot}
\end{figure*}
\begin{figure*}[!hbtp]
\centering
\includegraphics[trim=0.cm 4.cm 0.cm 4.cm,width=0.9\textwidth]{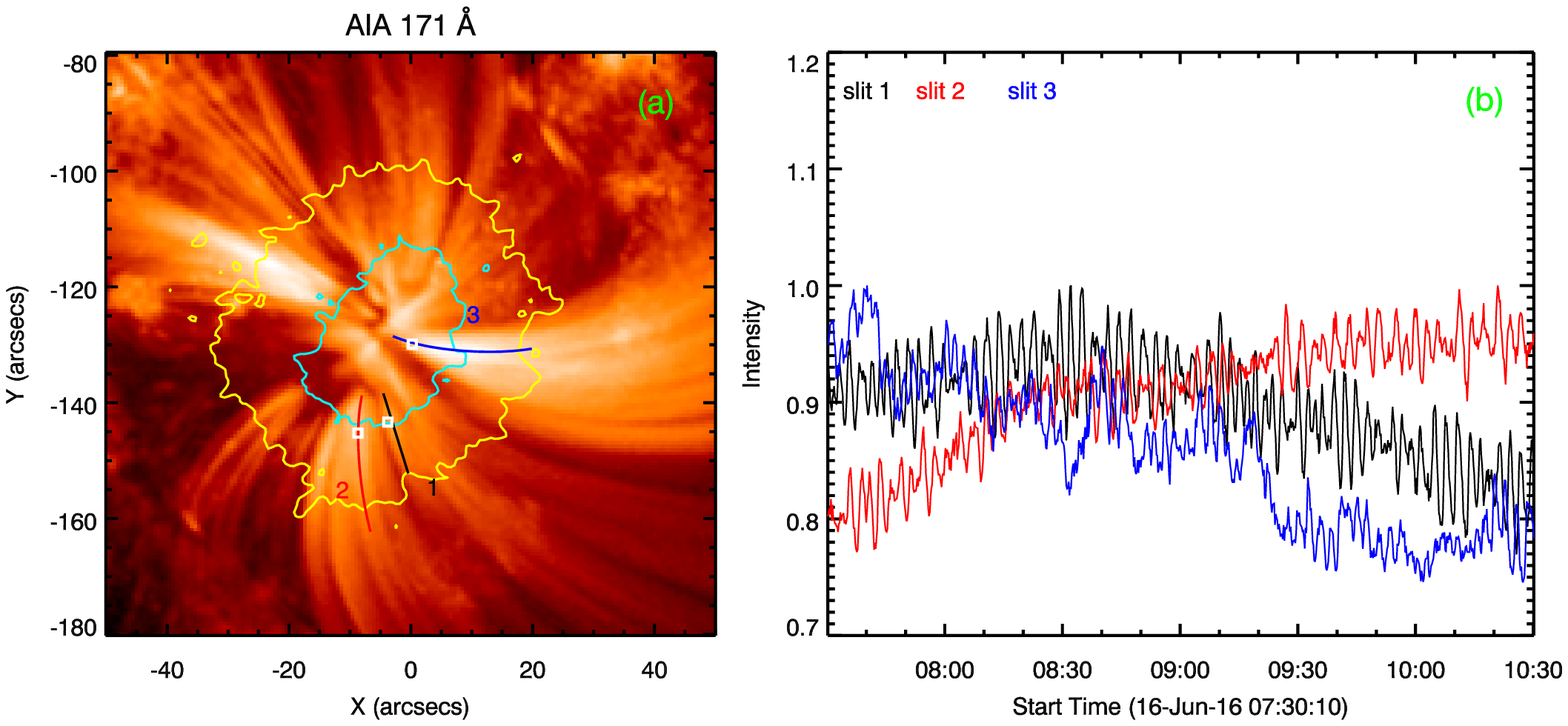}
\caption{Panel (a) Fan loop system traced with black, red and blue curved slits to perform time-distance analysis. The small white boxes locate the region, which is considered to study the steadyness of the fan loops. The contours are umbra-penumbra and penumbra-outer sunspot boundaries as described in Fig.~\ref{fig:sunspot}. Panel (b) Normalised light curves in AIA 171~{\AA} obtained for the white boxes shown in panel (a).}
\label{fig:lc}
\end{figure*}

The fan loop system, which is studied here, belongs to the active region NOAA \textit{AR 12553} observed on June 16, 2016. We have analysed the AIA observations recorded between 07:30:10 to 10:30:10~UT. We note that all the observations taken in different channels are rotated/de-rotated to the reference time at 9:15:00 UT for further analysis. Panel (a) in Fig.~\ref{fig:ss171} displays a portion of the Sun's disk image including \textit{AR 12553} observed in 171~{\AA} passband. The yellow box encloses the fan loop system named as region of interest, a zoom-up of which is shown in panel (b). 
\section{Data Analysis and Results} \label{analysis}
In Fig.~\ref{fig:sunspot}, we plot the regions of interest observed from HMI and AIA. Panels (a) and (b) show the HMI intensity and line-of-sight [LOS] magnetogram, respectively. The magnetogram is displayed within the magnetic field range of $\pm 1000$ G (with white indicating positive polarity and black indicating negative polarity). Panel (c) to (i) are taken by AIA in passband as labelled. The over-plotted cyan inner and yellow outer contours are plotted to visualise the umbra-penumbra and penumbra-plage boundaries, respectively. The umbra-penumbra contour is at the level of 38000~DN, whereas the penumbra-plage boundary is at the level of 53000~DN based on HMI continuum image. These contours are further over-plotted on the UV and EUV intensity images obtained in 1600~{\AA} and 171~{\AA}, shown in panels (c) and (e). Panel (e) reveals that some fan loops are rooted deep inside the umbra, and some are at the umbra-penumbra boundary. 

Since our primary motivation is to study the wave dynamics of steady fan loops, we focus on the region that does not suffer from any flaring activity. With this condition in mind, we located three different regions on the fan loop system with white boxes, as shown in the panel (a) of Fig.~\ref{fig:lc} and studied the corresponding light curves over the following three-hour period. The resultant light curves are shown in panel (b) of Fig.\ref{fig:lc}. The light curves are normalized with respect to the maximum intensities of each curve. The light curves conspicuously give the impression that the fan loop system does not show evolution more than 10\% from the mean. Therefore, these could be considered for our study. The blue light curve taken from a location along fan loop 3 indicates some background intensity enhancement from time to time.

In order to study the existence and propagation of intensity disturbances along the fan loops, we employ time-distance analysis and construct artificial slits (slits 1, 2 and 3) along the three fan loops as shown in panel (a) of Fig.~\ref{fig:lc}. We choose 171~{\AA} image for the identification of fan loops as they are best seen at temperature of $\log\,[\textit{T/K}] = 5.80$ \citep [e.g.,][]{Brooksetal2011, 2017ApJ...835..244G}. We note that each slit is four pixels wide. We further emphasise that this analysis could be performed using one slit. However, to check the consistency of the results, we have used three different slits along different fan loops anchored in the umbra of the sunspot (see panel (a) in Fig.~\ref{fig:lc}). We describe the results for slit 1 in the following, and provide the results for slits 2 and 3 in the Appendix for brevity.
\subsection{Time-distance maps}
\begin{figure*}[htbp]
\centering
\includegraphics[width=0.75\textwidth]{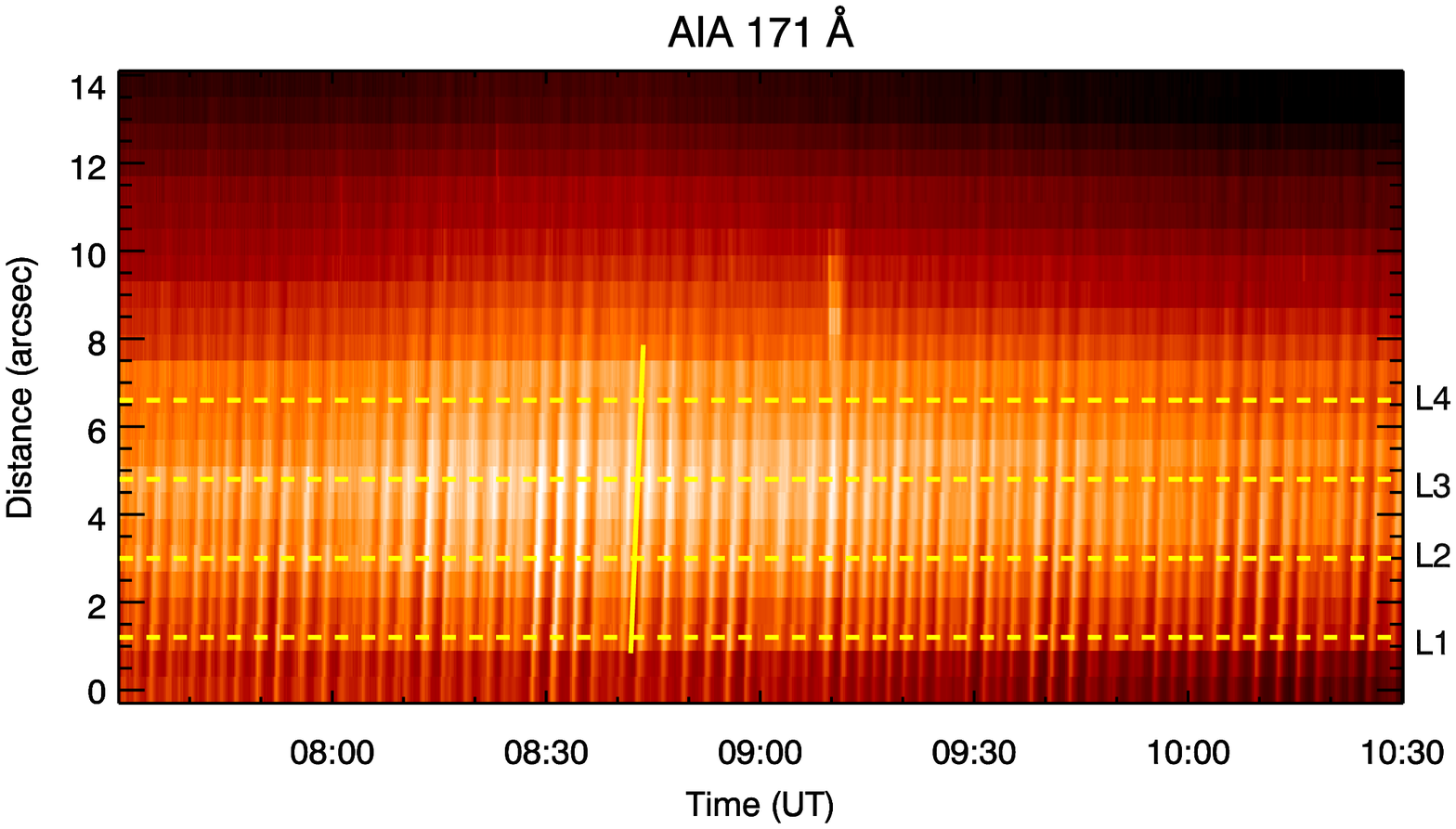}
\caption{Time-distance map obtained along artificial slit 1 using 171~{\AA} running difference images. The yellow horizontal dashed lines show the locations considered for obtaining the light curves for further analysis. The slanted yellow line along the propagating feature is used to estimate average projected speed.}  
\label{fig:xt171}
\end{figure*}
\begin{figure}[!hbtp]
\centering
\includegraphics[trim=1.0cm 0.0cm 0.0cm 0.0cm,width=0.5\textwidth]{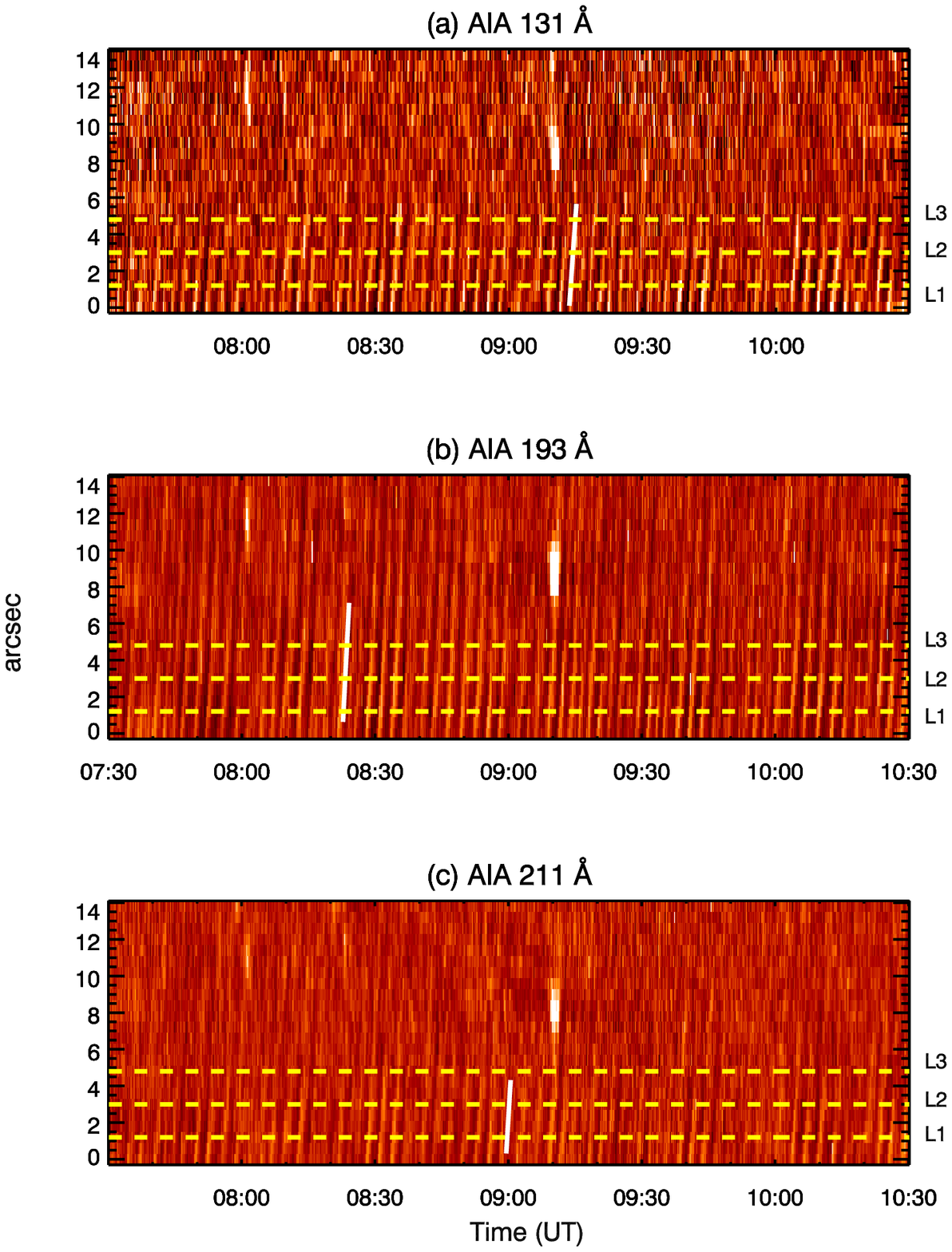}
\caption{Time-distance maps obtained along artificial slit 1 in different AIA EUV channels as labelled. The yellow horizontal dashed lines mark the locations chosen to obtain the light curves for further analysis. The white slanted lines are to obtain average projected wave speeds as mentioned in Fig.~\ref{fig:xt171}.} 
\label{fig:xts1_all}
\end{figure}
Fig.~\ref{fig:xt171} displays the time-distance map observed along slit 1 in 171~{\AA}. 
The time-distance map reveals clear propagation of intensity disturbances in the form of dark and bright ridges. These ridges are observed up to a distance of 8-10$\arcsec$\ (5.90-7.35 Mm) along the slit.
 
Fig.~\ref{fig:xts1_all} displays the time-distance maps obtained  along slit 1 from other AIA EUV channels as labelled. The PIDs are clearly discernible in all the EUV channels of AIA except for channels like 335~{\AA}, and 94~{\AA}, which are formed above a temperature of 2.5 MK. In AIA 193~{\AA}, the PIDs are observed up to a distance of 8-10$\arcsec$\ (5.9-7.35 Mm) similar to 171~{\AA} channel of AIA. However, in channels such as 131~{\AA}, and 211~{\AA} the PIDs are only seen upto a distance of around 4$\arcsec$\ (2.9 Mm).
The time-distance maps also reveal a decrease in the signal strength of the propagating features as 
they travel along the loops, which indicates a reduction in their amplitude of oscillations ($\approx$ 10\% to 5\%). 

\begin{figure*}[!hbtp]
\centering
\includegraphics[width=0.8\textwidth]{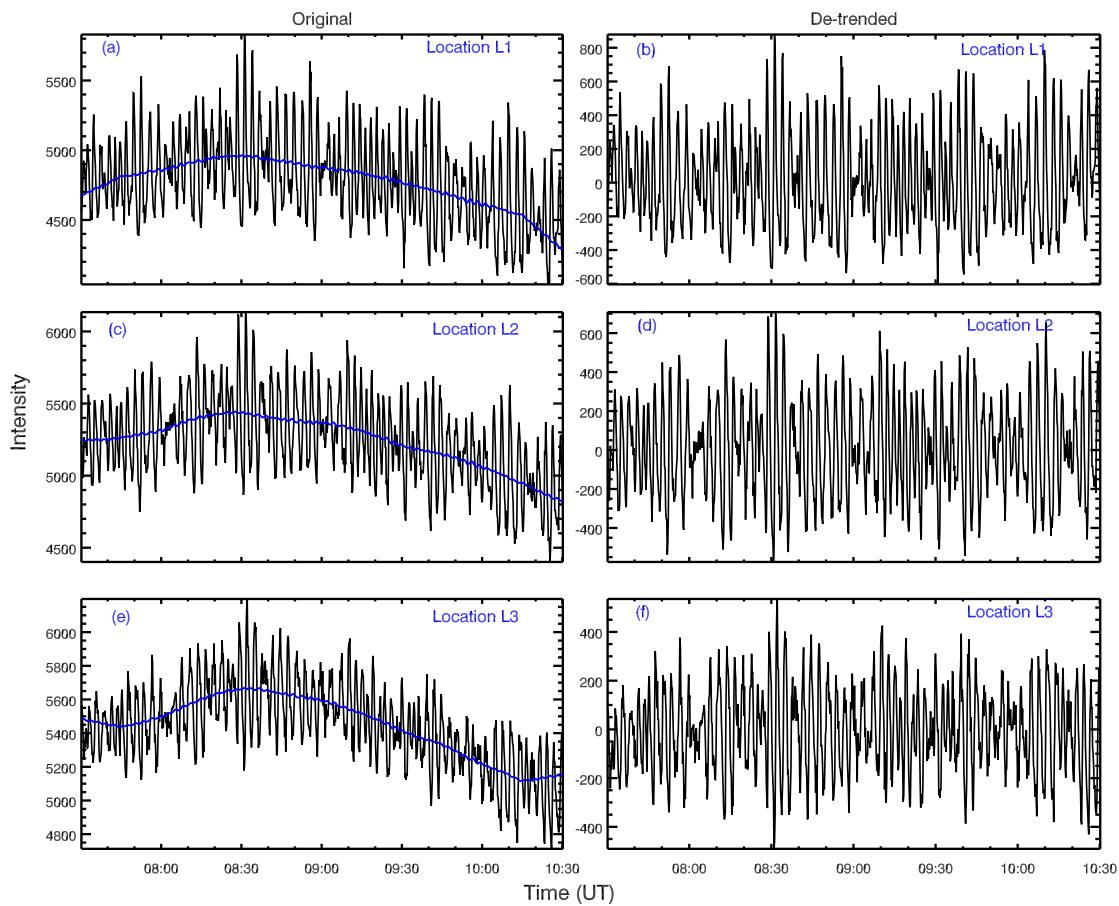}
\caption{Original (left column) and de-trended (right column) light curves obtained in AIA 171~{\AA} from different locations as marked along slit 1. The overplotted blue lines on the left panels mark the background trend of respective locations. }
\label{fig:or_dtred_lc}
\end{figure*}

To perform a quantitative analysis on the propagation of the intensity disturbances, we obtain light curves at different projected heights from the origin of the slit. The yellow dashed lines in Figs.~\ref{fig:xt171} and \ref{fig:xts1_all} highlight the locations chosen to receive the light curves. We denote the locations by L1, L2, L3, and L4 that correspond to $1.2\arcsec$, $3.0\arcsec$, $4.8\arcsec$, and $6.6\arcsec$, respectively. We note that the light curves are obtained from the absolute, unprocessed time-distance maps. We show such light curves in 171~{\AA} for a few projected heights in Fig.~\ref{fig:or_dtred_lc} (left column). In the right column of Fig.~\ref{fig:or_dtred_lc}, we plot corresponding de-trended light curves. The over-plotted blue lines on the original light curves indicate the background trends, which are obtained from a running average over 30 minutes. We see that 30-min running average describe the low-frequency background very well. We then, subtract the background trends from the respective original light curves to obtain the de-trended light curves. The de-trended light curves provide a very clear, evident, and unperturbed evolution of the intensity. We use the de-trended light curves for further studies.
\subsection{Phase speeds of the Propagating Intensity Oscillations}\label{velocity}
\begin{figure}[!hbtp]
\centering
\includegraphics[width=0.45\textwidth]{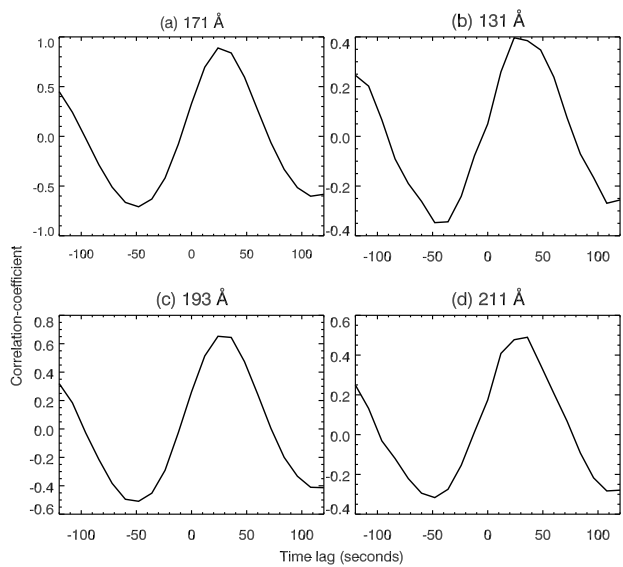}
\caption{Cross-correlation analysis performed between de-trended light curves obtained from locations L1 and L2 of the time-distance maps (locations shown by yellow dashed lines in Figs.~\ref{fig:xt171} and \ref{fig:xts1_all}). The cross-correlation coefficient is plotted on \textit{y}-axis as a function of time lag in seconds on \textit{x}-axis.}
\label{fig:cross_corr_allwav}
\end{figure}
\begin{table*}[htbp]
\renewcommand\thetable{1} 
\centering
\caption{\label{tab1} Phase speeds of the PIDs in different EUV channels obtained from cross-correlation analysis. Phase speeds obtained using the ridges of time-distance measurements are also presented for better comparison.}
\resizebox{14cm}{!} {
\begin{tabular}{cccc}
\space
\space
\hline
\hline
{AIA EUV channel}	&{Time lag}         &{Phase speed from cross-correlation analysis}   &{Phase speed from time-distance ridges}	  \\   

{{\AA}}        &{sec.}	             &km~s$^{-1}$	 &km~s$^{-1}$  \\    
\hline
\hline

                &                     &  	 &  	     \\
171             &    24.3               & 54.44      & 50.0     \\
131             &    24.4               & 54.22     & 50.0     \\ 
193             &    24.4              & 54.22      & 57.7    \\ 
211            &    35.55              & 37.21        & 49.0    \\

\hline
\hline
\end{tabular}
}
\label{table1}
\end{table*}
Before going into the details of the analysis, we intend to characterize the PIDs in terms of their phase speeds, as its measurements play a vital role in understanding their nature. To obtain the phase speeds, we employ two methods. In one method, we obtain the phase speeds by obtaining the slopes of the ridges observed in the time-distance plots. We draw multiple slopes (more than 20) along the propagating ridges by visual inspection and obtained the average propagation speed. The over-plotted yellow and white slanted lines in Figs.~\ref{fig:xt171} and \ref{fig:xts1_all} indicate the slopes used to measure the average projected phase speeds of the PIDs. We obtain the average projected phase speeds along slit 1 as $50.0\pm 9.0$~km~s$^{-1}$ for 171~{\AA}, $50.0\pm 10.0$~km~s$^{-1}$ for 131~{\AA}, $57.7\pm 10.0$~km~s$^{-1}$ for 193~{\AA}, and $49.0\pm 13.6$~km~s$^{-1}$ for 211~{\AA}. The noted errors are the standard deviation obtained using multiple slopes. 

Although the phase speeds obtained above by manually drawing slopes provides a feasible estimate of the phase speeds, it is subjected to errors due to its user-dependent nature. Therefore, in order to avoid the human bias associated with this method, we employ a technique that provides phase speeds by cross-correlating the light curves obtained at different heights \citep[e.g.,][]{2010ApJ...718...11G, 2012SoPh..279..427K}. The cross-correlation yields the time lag corresponding to the maximum cross-correlation coefficient and thereby the phase speeds. We carry out this analysis by using the standard IDL routine $c\_correlate$. For this purpose, we consider the de-trended light curves obtained at locations L1 and L2 (marked by yellow dashed lines in Figs.~\ref{fig:xt171} and \ref{fig:xts1_all}). We emphasise that the analysis carried out with the original light curves also gives similar time lags.

The cross-correlation curves obtained for different AIA EUV channels are shown in Fig.~\ref{fig:cross_corr_allwav}. 
The obtained phase speeds are given in Table~\ref{table1} against different time-lags observed in different EUV channels. For comparison, we have also listed the phase speeds achieved using the ridges of time-distance maps. We note that the obtained phase speeds using both the methods are in agreement within the errors. 

The obtained phase speeds are not the true speed but projected speeds in the plane of sky, as mentioned earlier. In order to measure the true phase speed, we need an estimate of the inclination angle (the angle between the line of sight and the guided magnetic field along which the wave propagates). The true phase speed is then obtained from the measured projected phase speed divided by the sine of the inclination angle. Since we do not have direct measurement of the inclination angle of the guide field, we consider a range of possible angles varying between $10\degree$ to $90\degree$ and estimate the respective true phase speeds for different AIA channels. Our findings show that the true phase speeds remain close to the sound speed in corona\footnote{The speed of sound in the solar corona is given by  $0.14T^{1/2}$~km~s$^{-1}$ with $T$ measured in kelvin [K] \citep[see][]{2004psci.book.....A}}. The sound speed for 1 MK coronal plasma is around 140~km~s$^{-1} $. As an example, we discuss the estimated true phase speeds of the PIDs in AIA 171 ~{\AA} channel considering different inclination angles in the Appendix A. The phase speeds of the PIDs closer to the speed of sound suggest that they are slow mode magneto-acoustic waves.
\subsection{Amplitude modulation observed in AIA 171~{\AA}}\label{lightcurve}
The de-trended light curves in 171~{\AA} (Fig.~\ref{fig:or_dtred_lc}) reveal a peculiar oscillatory pattern. We find the clear presence of an amplitude modulation with an overall increase and decrease of the wave pattern over time. The modulation is discernible even in the time-distance map shown in Fig.~\ref{fig:xt171} in terms of brightness.

In order to understand the modulation in more details, we have obtained the frequency components responsible for such patterns using Fouri\'er analysis and studied their evolution using wavelet analysis \citep{1998BAMS...79...61T}.
\subsubsection{Fouri\'er Analysis}\label{ft171}
\begin{figure*}[!hbtp]
\centering
\includegraphics[width=12.cm]{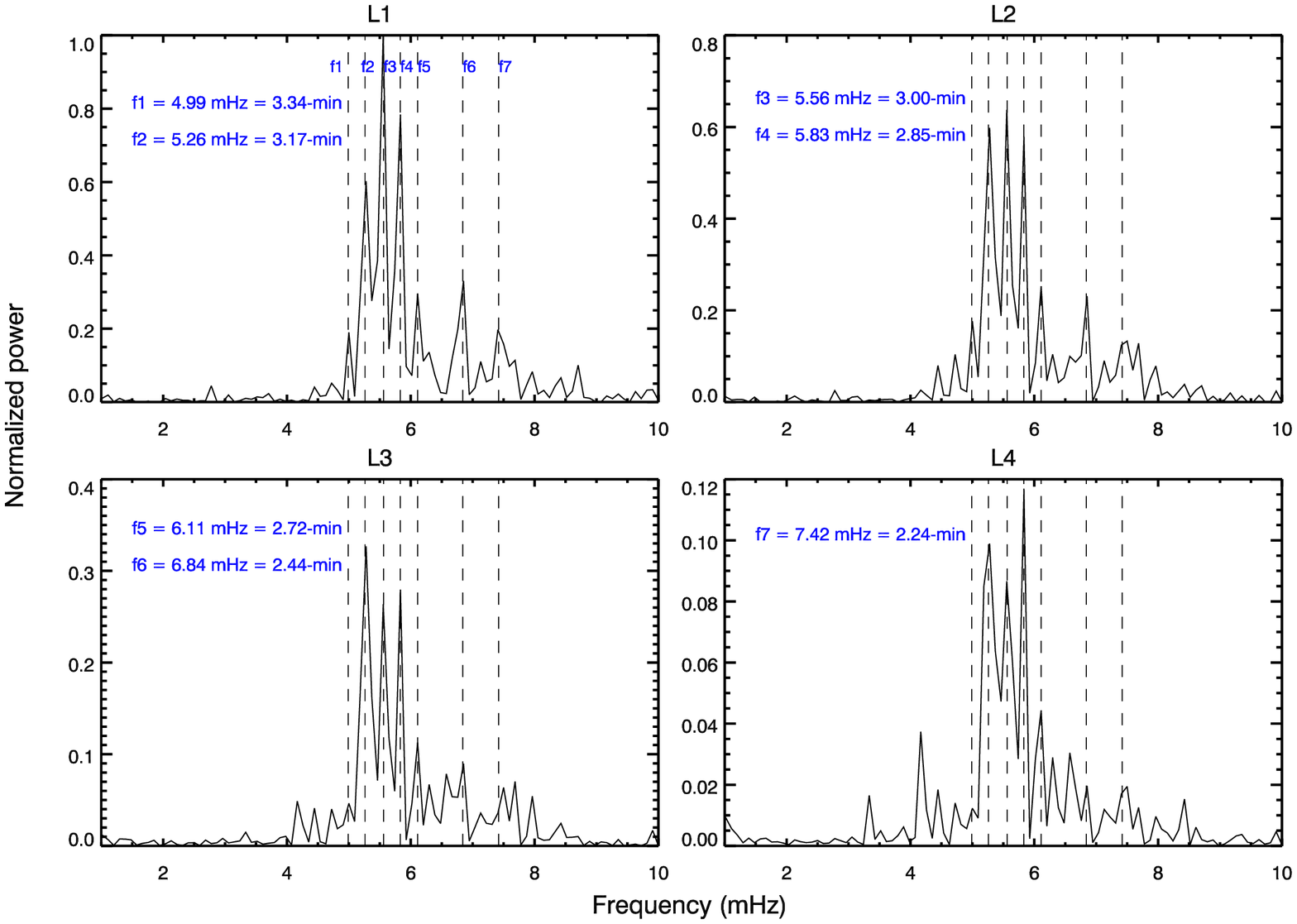} 
\caption{Fouri\'er power spectrum of the original light curves obtained from different locations of AIA 171~{\AA} time-distance map (locations marked in Fig.~\ref{fig:xt171}).} 
\label{fig:fft171}
\end{figure*}
Fig.~\ref{fig:fft171} displays the normalised Fouri\'er power spectra derived from the original light curves for the locations L1 -- L4 in Fig.~\ref{fig:xt171} (marked by yellow dashed lines).
The power in the panels is normalised with respect to the maximum power obtained at location L1. The power spectra of all the locations indicate three dominant peaks within 5 -- 6 mHz (3.33 -- 2.78 minutes). They are at around 5.26 mHz (3.17-min, marked as f2 in Fig.~\ref{fig:fft171}), 5.56 mHz (3.00-min, marked as f3), and 5.83 mHz (2.85-min, marked as f4). In addition, there are other smaller peaks identified at around 4.99 mHz (3.34-min, f1), 6.11 mHz (2.72-min, f5), 6.84 mHz (2.44-min, f6) and 7.42 mHz (2.24-min, f7). 
The vertical dashed black lines mark these peaks. We also observe a decrease in the power of the frequency peaks as we go away from the footpoint of the fan loop.

The observations of several nearby frequencies in the power spectra led us to postulate that the phenomenon of beat may play a role to generate the amplitude modulation. The beat is a phenomenon which occurs when two waves of slightly different frequencies, say, f$_1$ and f$_2$, interfere giving rise to a resulting wave undergoing amplitude modulations with frequency, f$_b$ = f$_1$-f$_2$. We call this frequency, ‘f$_b$’, as beat frequency. In the current observations, assuming the interaction of the three dominant frequency components (f2, f3, and f4) observed in Fig.~\ref{fig:fft171}, we find that for frequency components 5.26 mHz (f2) and 5.83 mHz (f4), there is a beat frequency f$_b$ = 0.57 mHz, i.e.,$\approx$ 29-min. We note that the 29-min is close to the modulation period that we obtain from a simple visual inspection of the detrended light curves in Fig.~\ref{fig:or_dtred_lc} (see also section \ref{wv171}).

The other combinations give us  f$_b$ = 0.30 mHz ($\approx$ 55-min) for 5.26 mHz (f2) and 5.56 mHz (f3) and f$_b$ = 0.27 mHz ($\approx$ 61.70-min) for 5.56 mHz (f3) and 5.83 mHz (f4). Similarly, the interference of the dominant frequency around 5.56 mHz (3-min.) with those of frequencies at around 4.99 mHz (f1), 6.11 mHz (f5), 6.84 mHz (f6), and 7.42 mHz (f7), give rise to beat frequencies of around 0.57 mHz (29-min), 0.55 mHz (30-min), 1.28 mHz (13-min), and 1.86 mHz (8.96-min), respectively. However, we note that in the dynamic and complex solar atmosphere, it is difficult to explain the observed period of amplitude modulation in terms of a particular beat period as the propagation becomes affected by all the frequency components. \citet{2015ApJ...812L..15K} observed amplitude modulation in the Fouri\'er-filtered light curves with a mean period of 20 -- 27 minutes, and reported the possibility of occurrence of the beat. 
We note that large-period oscillations with periods of around 8-min, 12-min, 24-min, 40-min, 63.4-min in coronal fan loops have also been reported before by \citet{2009A&A...503L..25W, 2011A&A...526A..58S}. However, no observational pieces of evidence were provided to answer their origin. 
\subsubsection{Wavelet Analysis}\label{wv171}
\begin{figure*}[htbp]
\centering 
\includegraphics[width=6.5cm,angle=90]{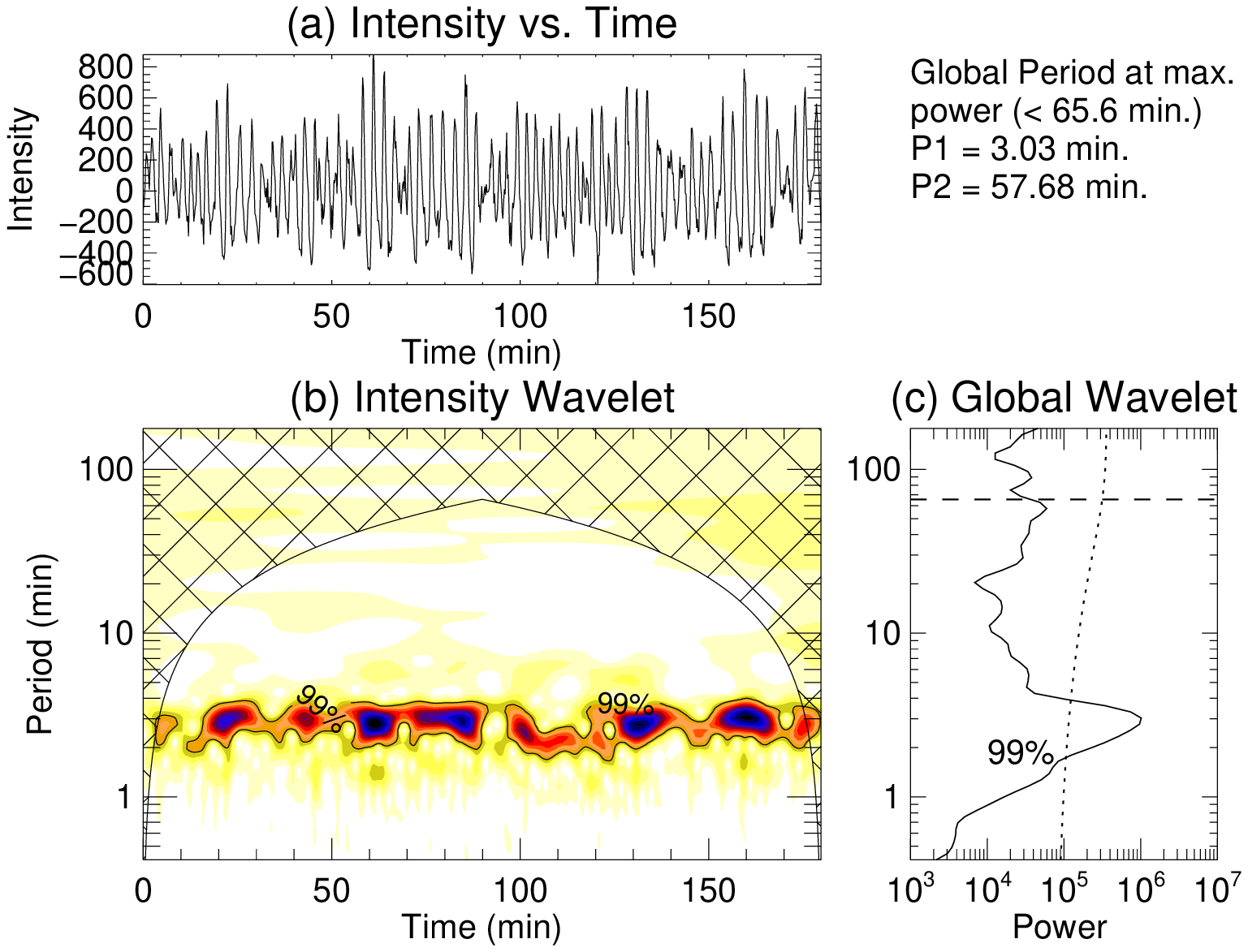}\includegraphics[width=6.5cm,angle=90]{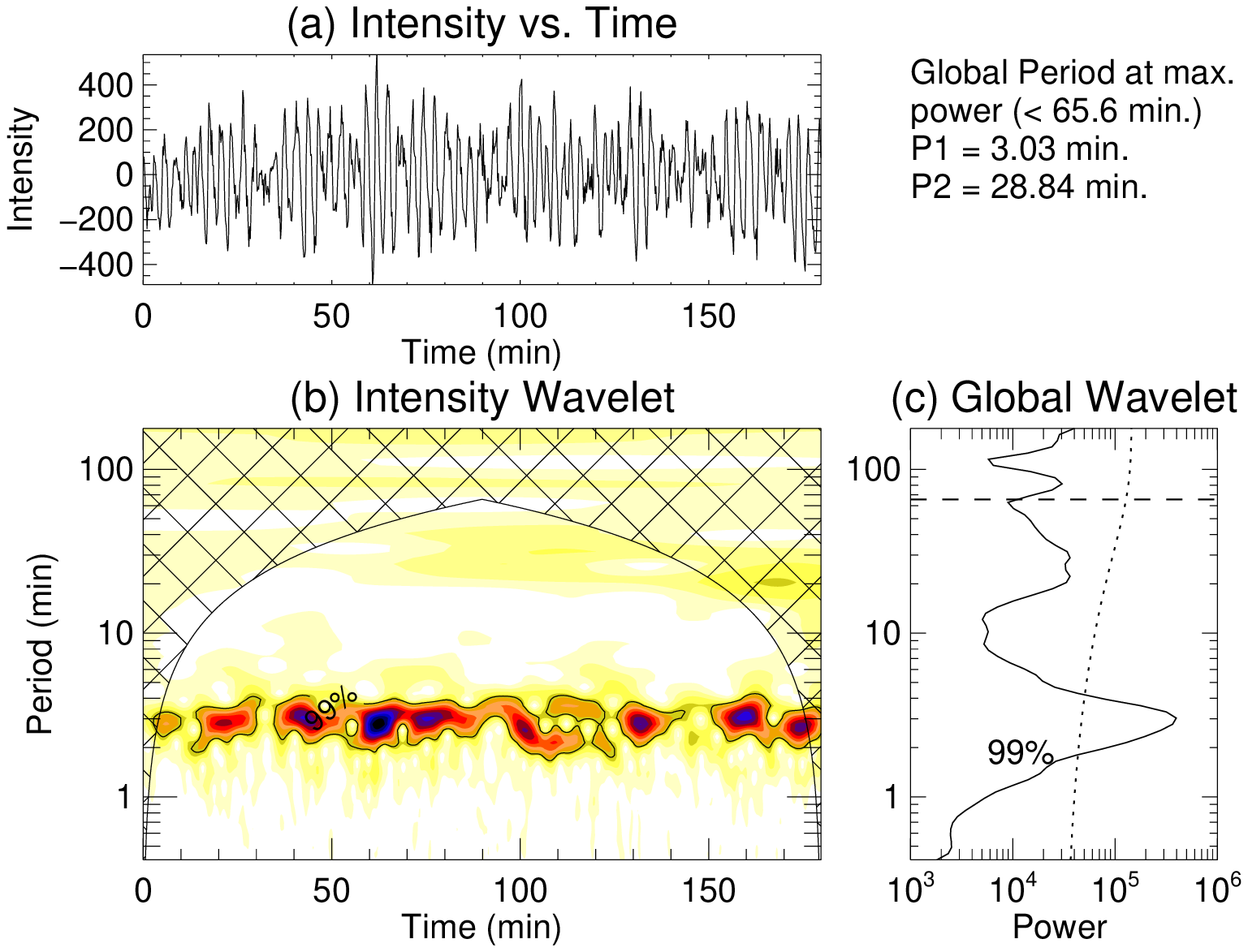}
\caption{Wavelet analysis results of de-trended light curves obtained from slit 1 AIA 171~{\AA} time-distance map for locations L1 (left) and L3 (right) (location marked in Fig.~\ref{fig:xt171}). The top panel is the variation of obtained de-trended intensity with time. The bottom left panel depicts the corresponding wavelet spectrum (blue shades for high power density), while the bottom right panel is the global wavelet power spectrum. Dashed lines in the global wavelet plots indicate the maximum period detectable from wavelet analysis due to cone-of-influence, whereas the dotted line indicates 99\% confidence level curve. Periods P1 and P2 of the first two global power peaks are printed at the top right. In the time-axis, the time starts at 8:00 UT.} 
\label{fig:wavelet171pix2_8}
\end{figure*}
\begin{figure*}[hbtp]
\centering 
\includegraphics[width=6.5cm,angle=90]{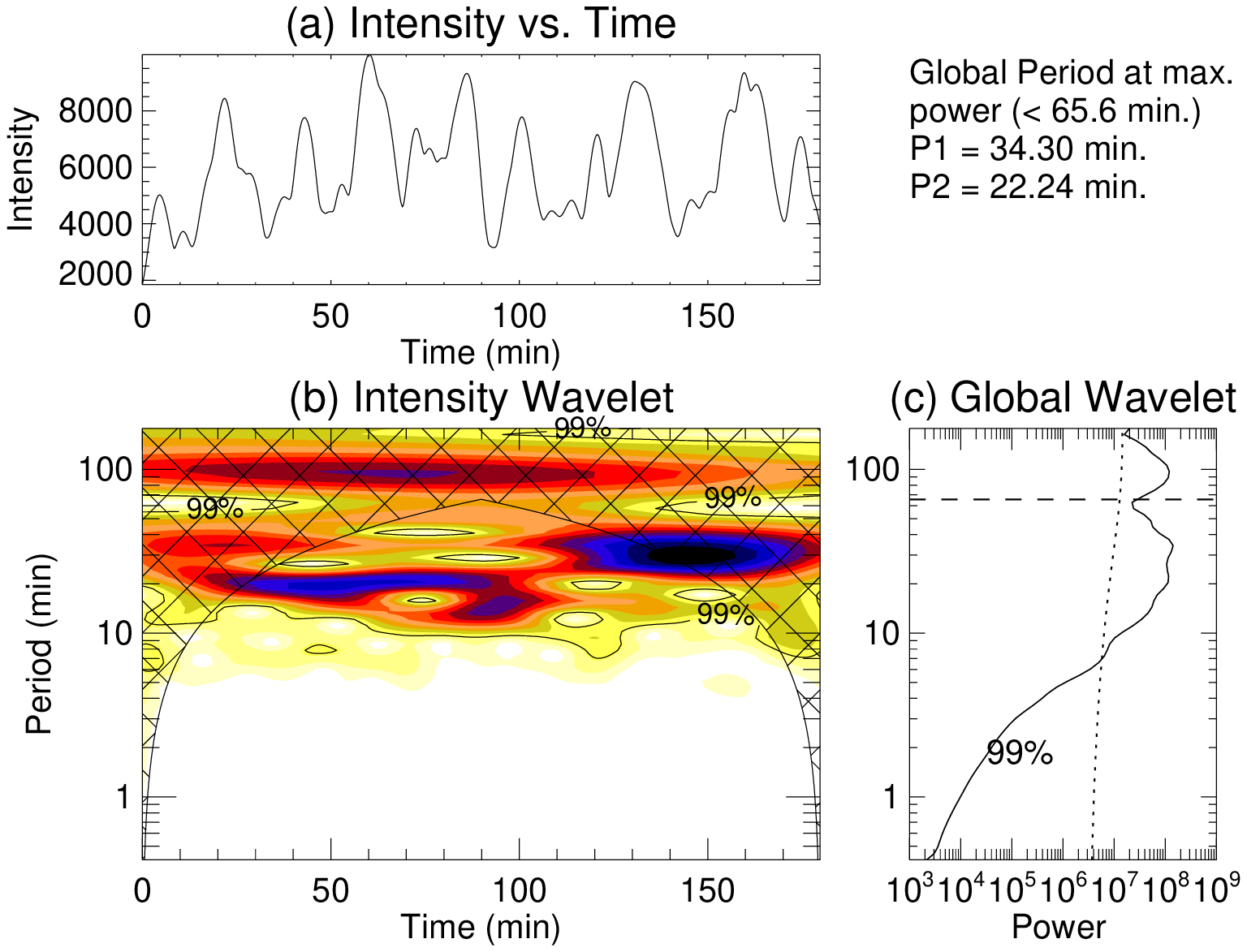}\includegraphics[width=6.5cm,angle=90]{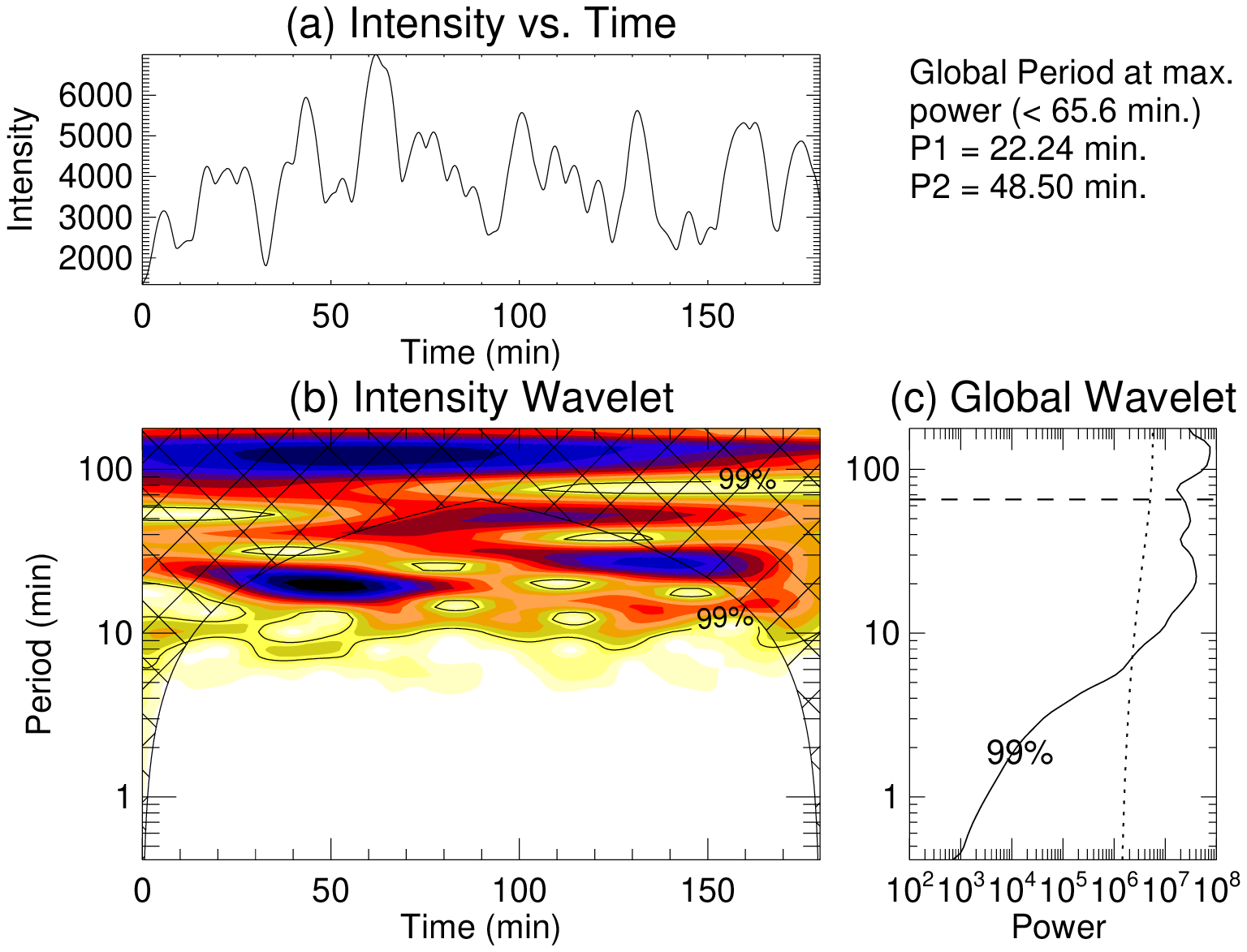}
\caption{Wavelet analysis result of 171~{\AA} light curves obtained by averaging waves with oscillation periods in the range 2-4 minutes. Wavelet plots are obtained for similar locations, L1 (left) and L3 (right) as mentioned in Fig.~\ref{fig:wavelet171pix2_8}.} 
\label{fig:wavelet_averageds1}
\end{figure*}
\begin{figure*}[!hbtp]
\centering 
\includegraphics[width=6.5cm,angle=90]{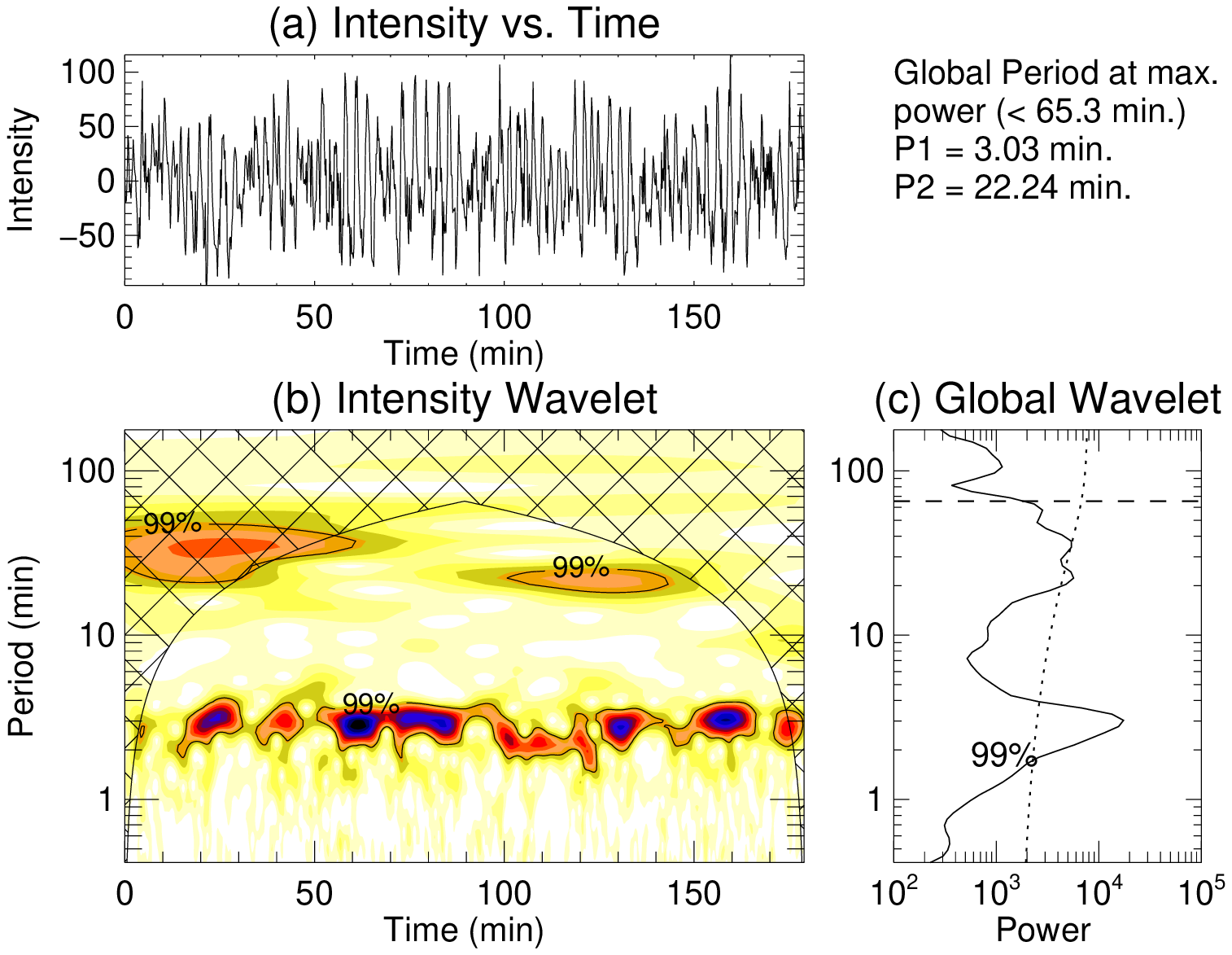}\includegraphics[width=6.5cm,angle=90]{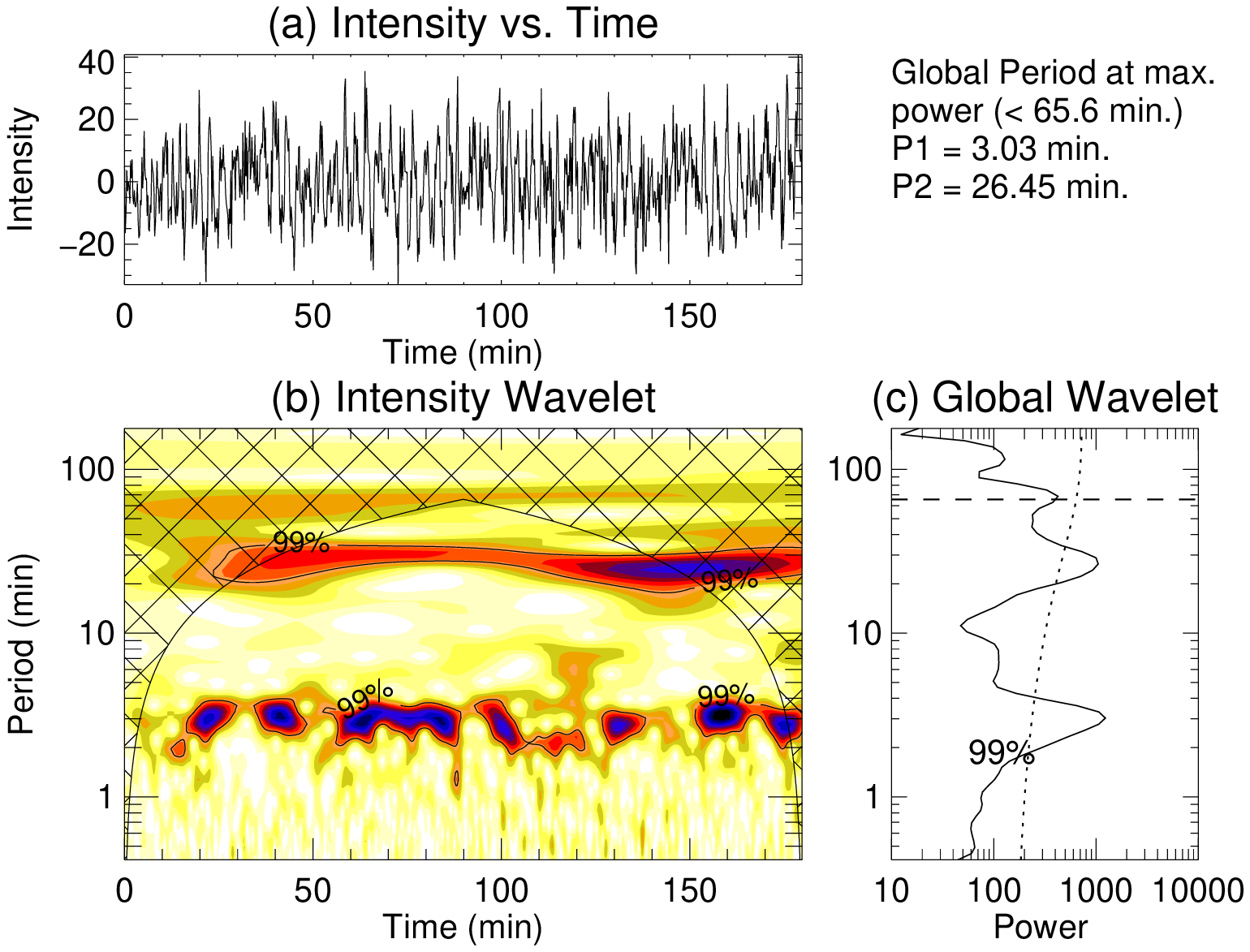}
\caption{Wavelet analysis results of de-trended light curves observed in AIA 193~{\AA} (left) and AIA 211~{\AA} (right) for location L1 of the time-distance maps (shown in Fig.~\ref{fig:xts1_all}). The panels are described in Fig.~\ref{fig:wavelet171pix2_8}.} 
\label{fig:wavelet193211}
\end{figure*}
\begin{figure}[!htbp]
\centering 
\includegraphics[width=0.36\textwidth]{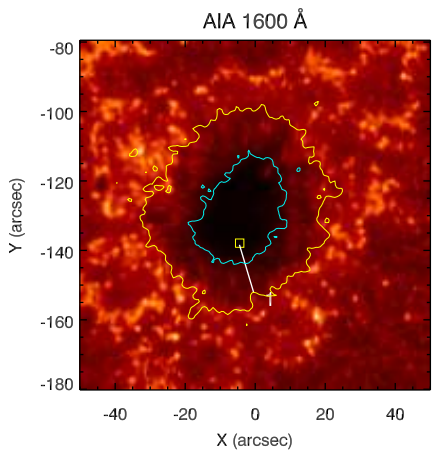}
\caption{Sunspot umbra in AIA 1600~{\AA}. The overplotted yellow box corresponds to the region at the footpoint of slit 1. The light curves averaged over the region are considered for further analysis in the chromospheric channels, e.g. AIA 304~{\AA} and 1600~{\AA}. Contours are similar to those described in panel (a) of Fig.~\ref{fig:lc}.} 
\label{fig:umbraloc}
\end{figure}
\begin{figure*}[!h]
\centering 
\includegraphics[width=5.5cm,angle=90]{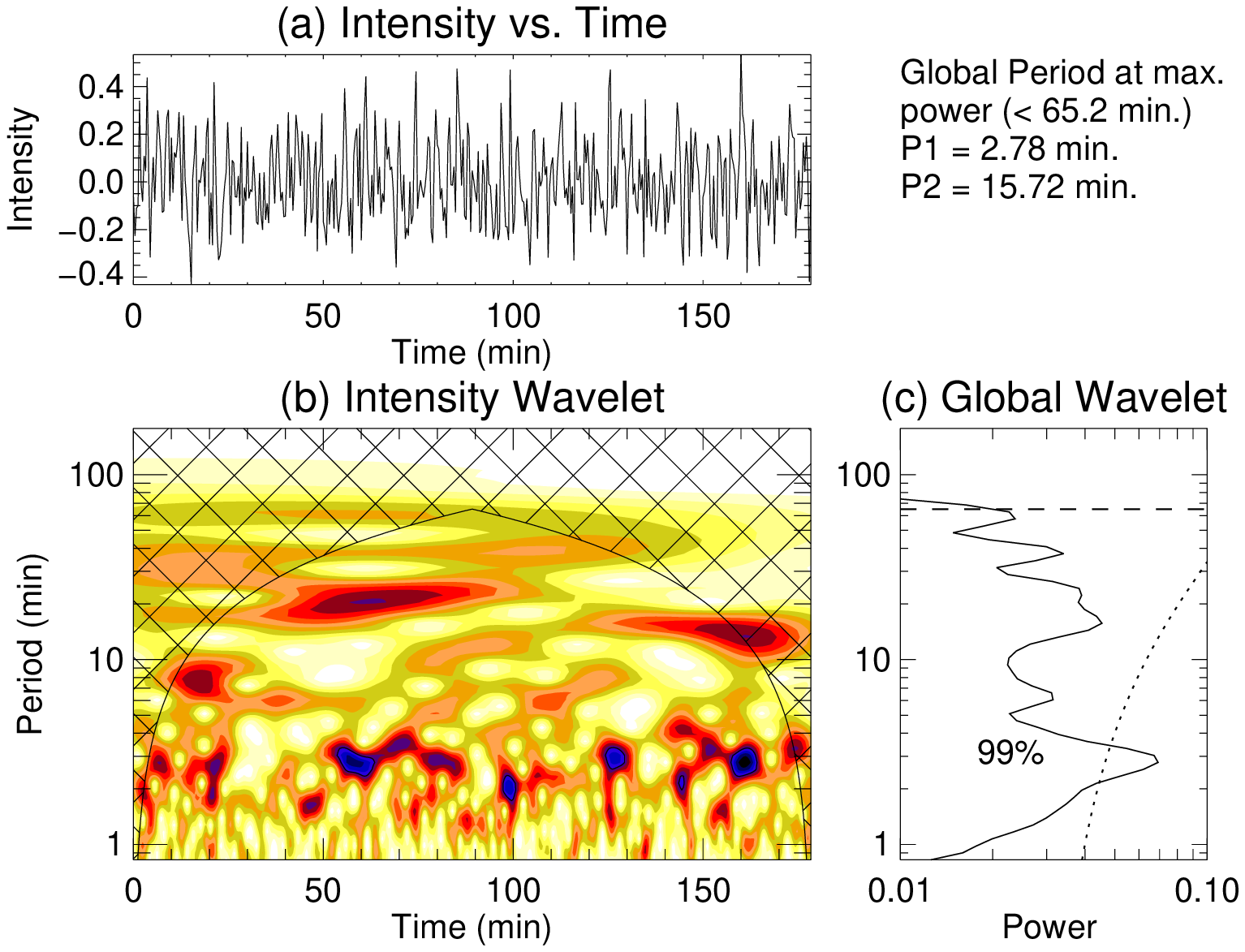}\includegraphics[width=5.5cm,angle=90]{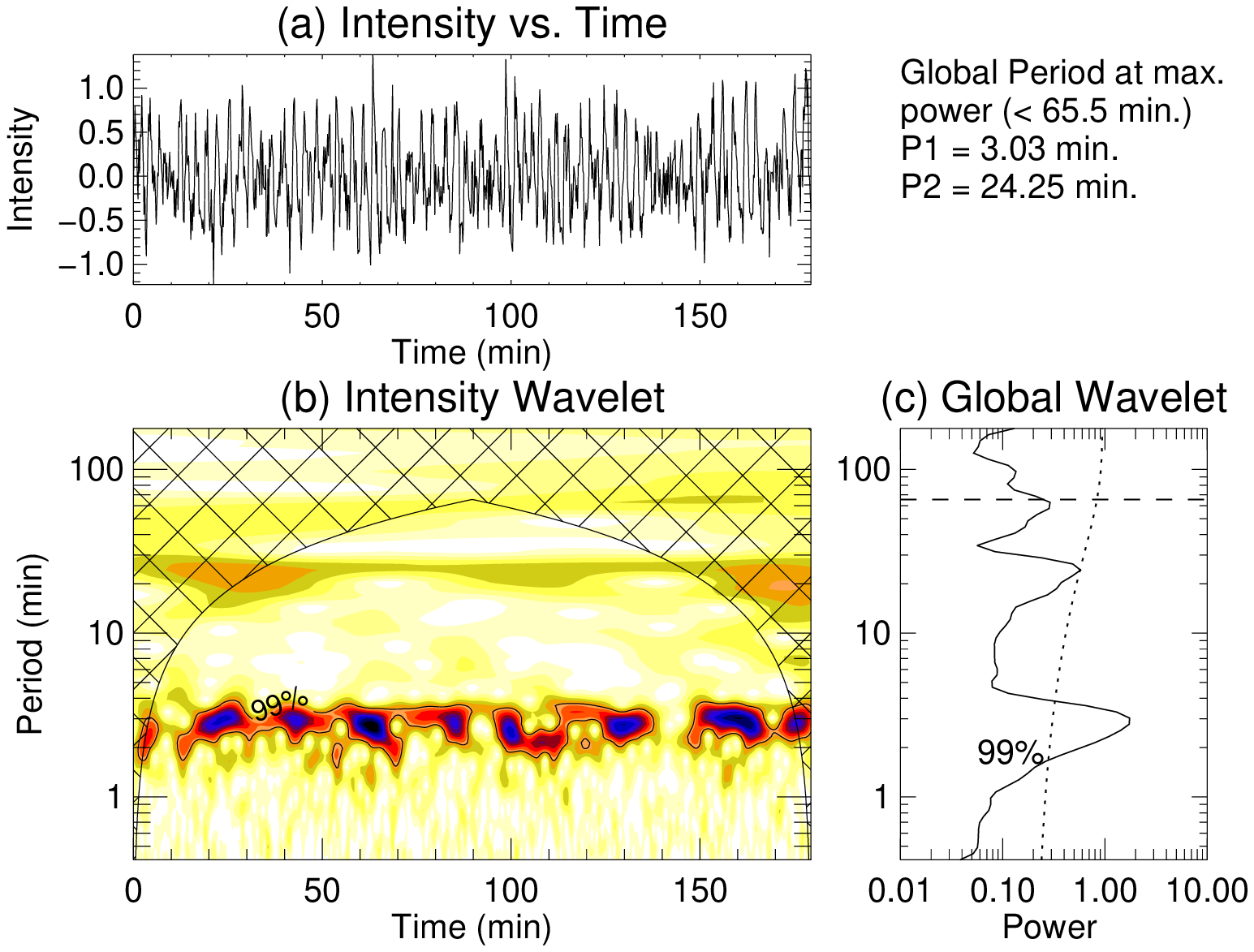}
\caption{Wavelet analysis of the de-trended light curves extracted from the yellow box location of Fig.~\ref{fig:umbraloc}. Left panels are obtained from 1600~{\AA}, and right panels are from 304~{\AA}. Different panels are explained in Fig.~\ref{fig:wavelet171pix2_8}. } 
\label{fig:waveletumbra}
\end{figure*}

To study the temporal variation of the frequency, we perform wavelet analysis on the 30-min de-trended 171~{\AA} light curves obtained at locations marked in Fig.~\ref{fig:xt171}. 
For this purpose, we use the Morlet wavelet as the basis function, which is a plane wave multiplied by a gaussian, and convolve it with the light curves \citep{1998BAMS...79...61T}. In Fig.~\ref{fig:wavelet171pix2_8}, we plot the wavelet results for the light curves obtained at locations, L1 (left panel) and L3 (right panel). Each wavelet plot consists of three panels. The de-trended light curve used for wavelet analysis is displayed on top panel (a). The wavelet power spectrum that shows the variation of power at different periods with time is plotted in the bottom left panel (b). The global wavelet power spectrum, obtained by taking the average over the time domain of the wavelet transform, is plotted in the bottom right panel (c). In each wavelet spectrum (lower left panel), we denote a cross-hatched region called the cone-of-influence (COI). The COI region appears because of the edge effects which arises due to the finite-length of time series. Any estimate of the oscillation period in this region becomes unreliable. The horizontal dashed lines in the global wavelet panels show the longest period that can be detected due to COI. 
We mention the first two power peaks (P1 and P2) obtained from the global wavelet at the right top corner of the wavelet plots.

In the wavelet spectrum panels, we find that most of the oscillating power lies in the range 2--4 minutes throughout the observed duration. We also observe an increase and decrease in the wavelet power, which is co-temporal with the appearance and disappearance of the modulations observed in the corresponding light curves. As expected, the global wavelet spectra show the presence of dominant power peak (P1) at around 3-min period for all the locations. We obtain second power peaks (P2) around 57.68-min. and 44.48-min. for locations L1 and L2, and at 28.84-min and 20.30-min for locations L3 and L4, respectively. We see that these second power peaks are below the 99\% confidence level, therefore unreliable.

To quantify the modulation period, we follow a particular approach \citep{2015ApJ...812L..15K, 2017ApJ...850..206S}. We see that it is the 3-min wave amplitude that is modulated. The Fouri\'er Power Spectrums (Fig.~\ref{fig:fft171}) also illustrate significant power within 2 to 4 minutes, with dominant peaks around 3-min. Inspecting the power within 2 to 4 minutes, we measure the wavelet oscillatory power averaged within this period range. This oscillatory power will essentially mimic an amplitude variation that would be caused by all the waves having periods within the 2 and 4-minute range \citep{2017ApJ...850..206S, 2015ApJ...812L..15K, 2012A&A...539A..23S}. Then, we apply wavelet to the obtained amplitude variation to estimate the dominant period of amplitude modulation. Fig.~\ref{fig:wavelet_averageds1} shows wavelet power spectrums obtained in such way for AIA 171~{\AA} for locations, as done for Fig.~\ref{fig:wavelet171pix2_8}. We obtain the dominant/first power peaks (P1) around 34.30-min, 20.39-min, 22.24-min, and 26.45-min for locations L1, L2, L3, and L4, respectively. We find the longer-period power peaks above the 99\% confidence level for all the marked locations of Fig.~\ref{fig:xt171}. We emphasise that we find similar results with the unfiltered, original light curves.
\subsection{Amplitude modulation in other AIA EUV channels}\label{othereuv}
For completeness, we study the light curves obtained from other AIA EUV channels at the locations shown by over-plotted yellow dashed lines of Fig.~\ref{fig:xts1_all}. We de-trended the light curves in a similar way as was done for those obtained from 171~{\AA} and performed further analysis. In Fig.~\ref{fig:wavelet193211}, we present the wavelet analysis results obtained using light curve at L1 for  193~{\AA} (left panels) and 211~{\AA} (right panels). The power is concentrated at a period (P1) around 3-min, and undergo an increase and decrease over time, as can be seen in the panels (b) of Fig.~\ref{fig:wavelet193211}. To quantify the modulation period, we approached the similar method as done for AIA 171~{\AA} (described in the last paragraph of \ref{wv171}). We obtain the dominant power peaks (P1) around 22.24-min, 20.39-min for 193~{\AA} and 211~{\AA} respectively The analysis of 131~{\AA} showed similar behaviour and provided the dominant modulation period around 22.24-min However, we emphasize that the modulation is best seen in AIA 171~{\AA}. 
\subsection{AIA 1600 ~{\AA} and 304 ~{\AA} observations at the footpoint of the fanloop}\label{footpoint}
To investigate the origin of the modulation shown by the propagating slow MHD waves in AIA EUV channels, we study the footpoint of the fan loop in AIA 1600~{\AA} and 304~{\AA}. Fig.~\ref{fig:umbraloc} shows an AIA 1600~{\AA} image with the footpoint of the fan loop obtained from AIA 171~{\AA} traced by slit 1. A yellow box bounds the footpoint location situated in the sunspot umbra. The size of the box is 4$\times$4 in pixel scale. We obtain the light curves averaged over the region and de-trend it over 30 minutes and performed wavelet analysis. The wavelet results for 1600~{\AA} (left panel) and 304~{\AA} (right panel) are shown in  Fig.~\ref{fig:waveletumbra}. In the wavelet power map, we do not observe a clear pattern in 1600~{\AA}, whereas in 304~{\AA}, there is clear modulation around 3-min. We obtain the modulation period around 22.24-min in the similar way as obtained for 171~{\AA} (described in last paragraph of \ref{wv171}).
\subsection{Observation of high frequency component}\label{hf}
\begin{figure*}[!hbtp]
\centering 
\includegraphics[width=0.37\textwidth]{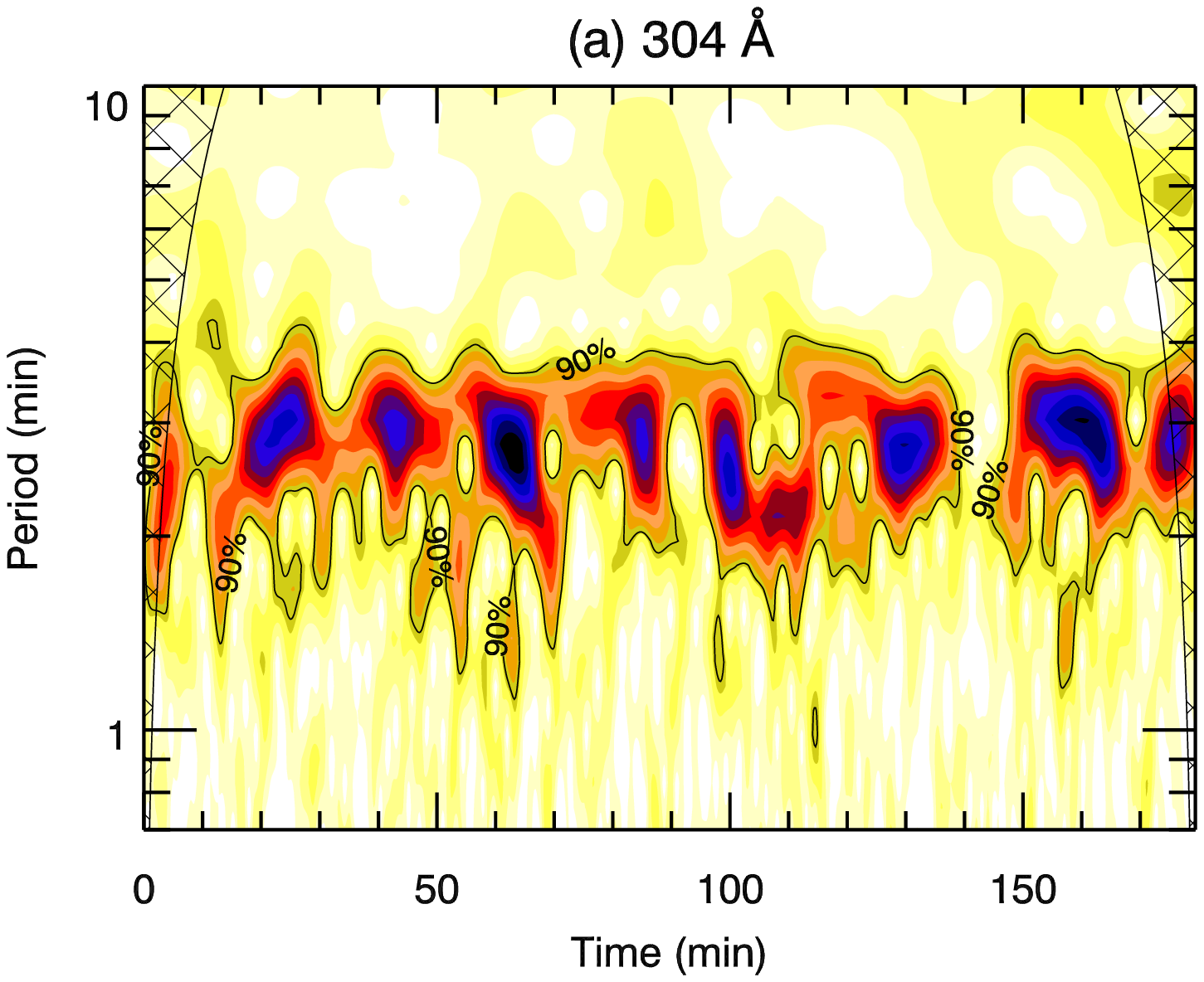}\includegraphics[width=0.37\textwidth]{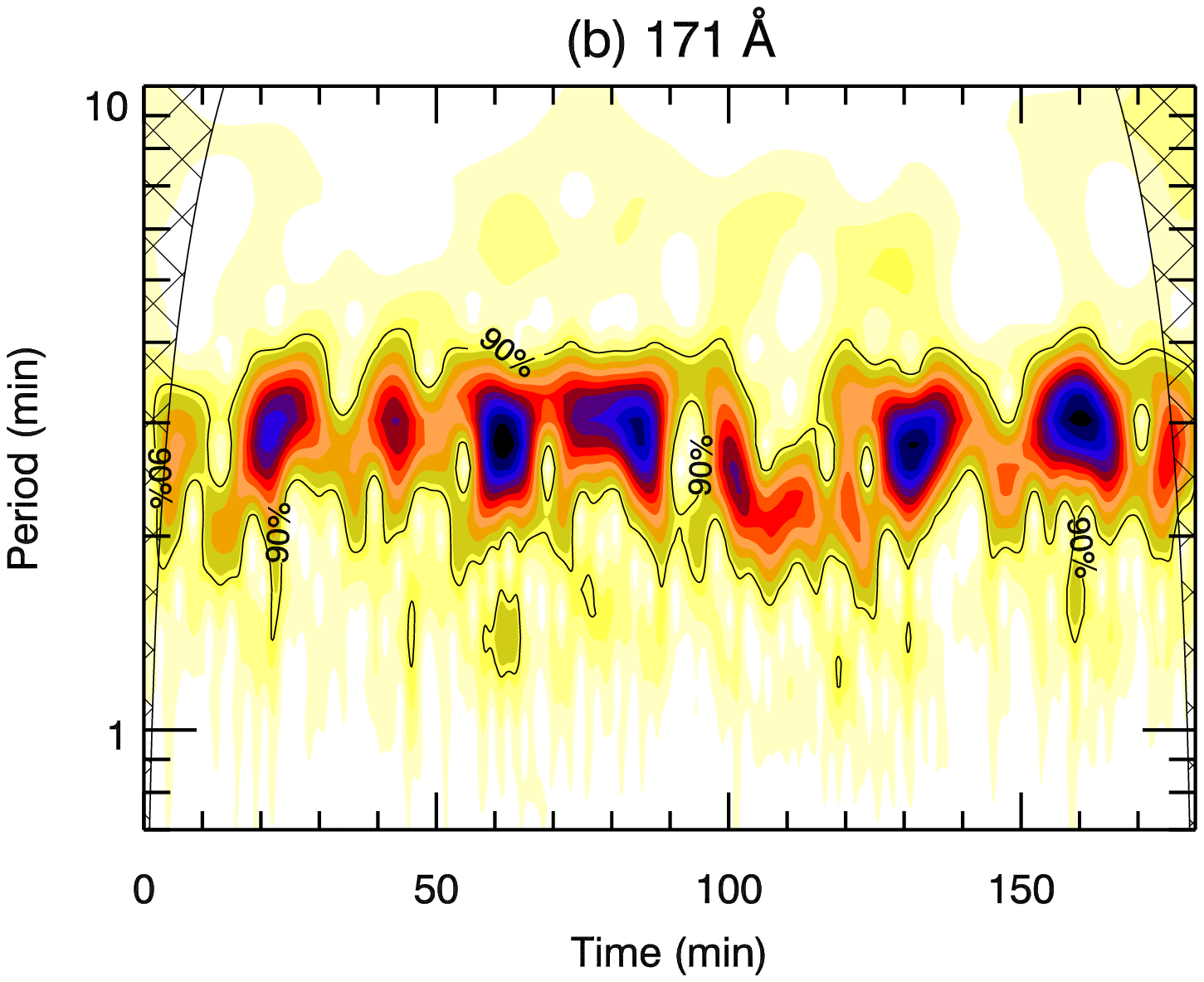}
\includegraphics[width=0.37\textwidth]{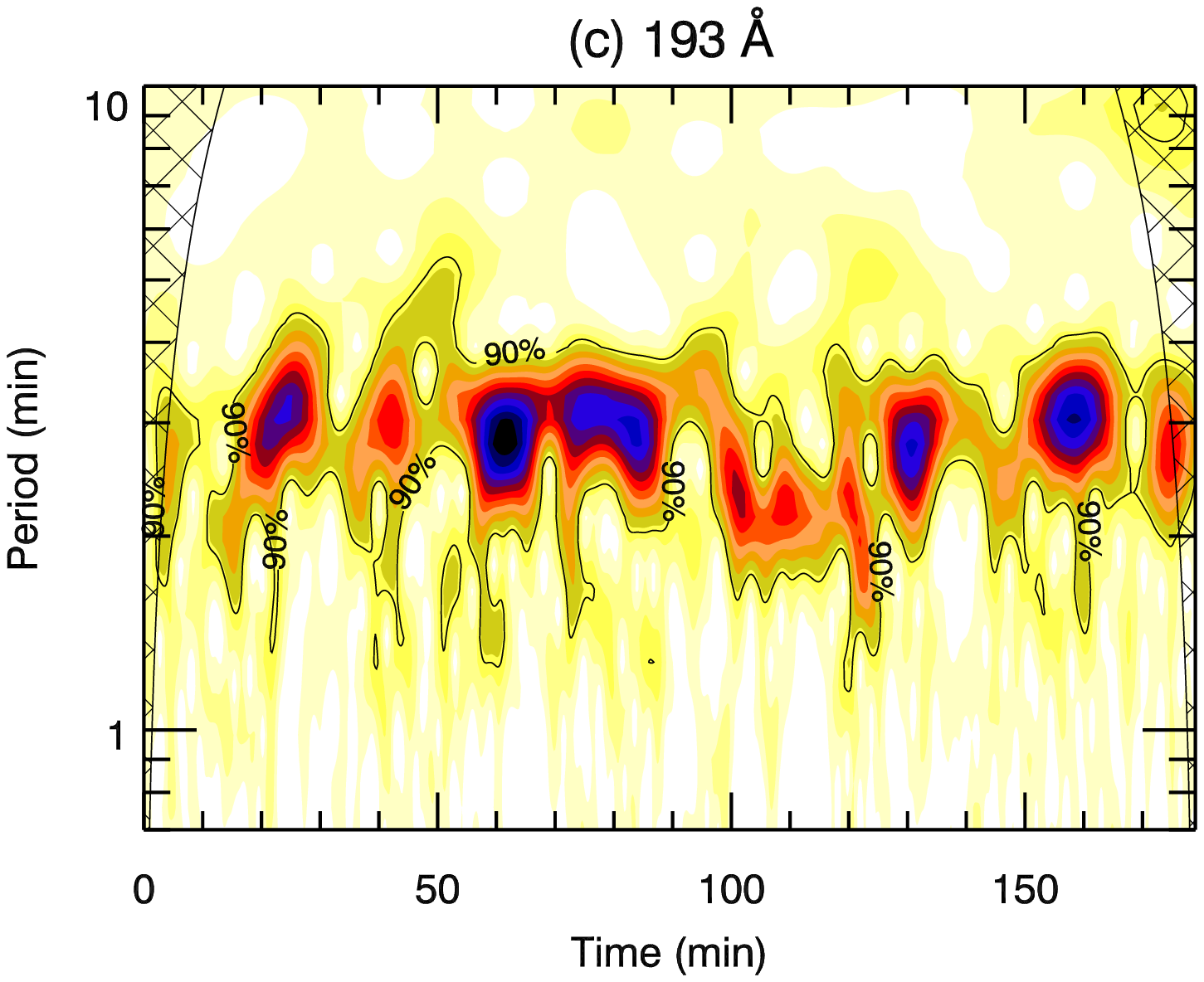}\includegraphics[width=0.37\textwidth]{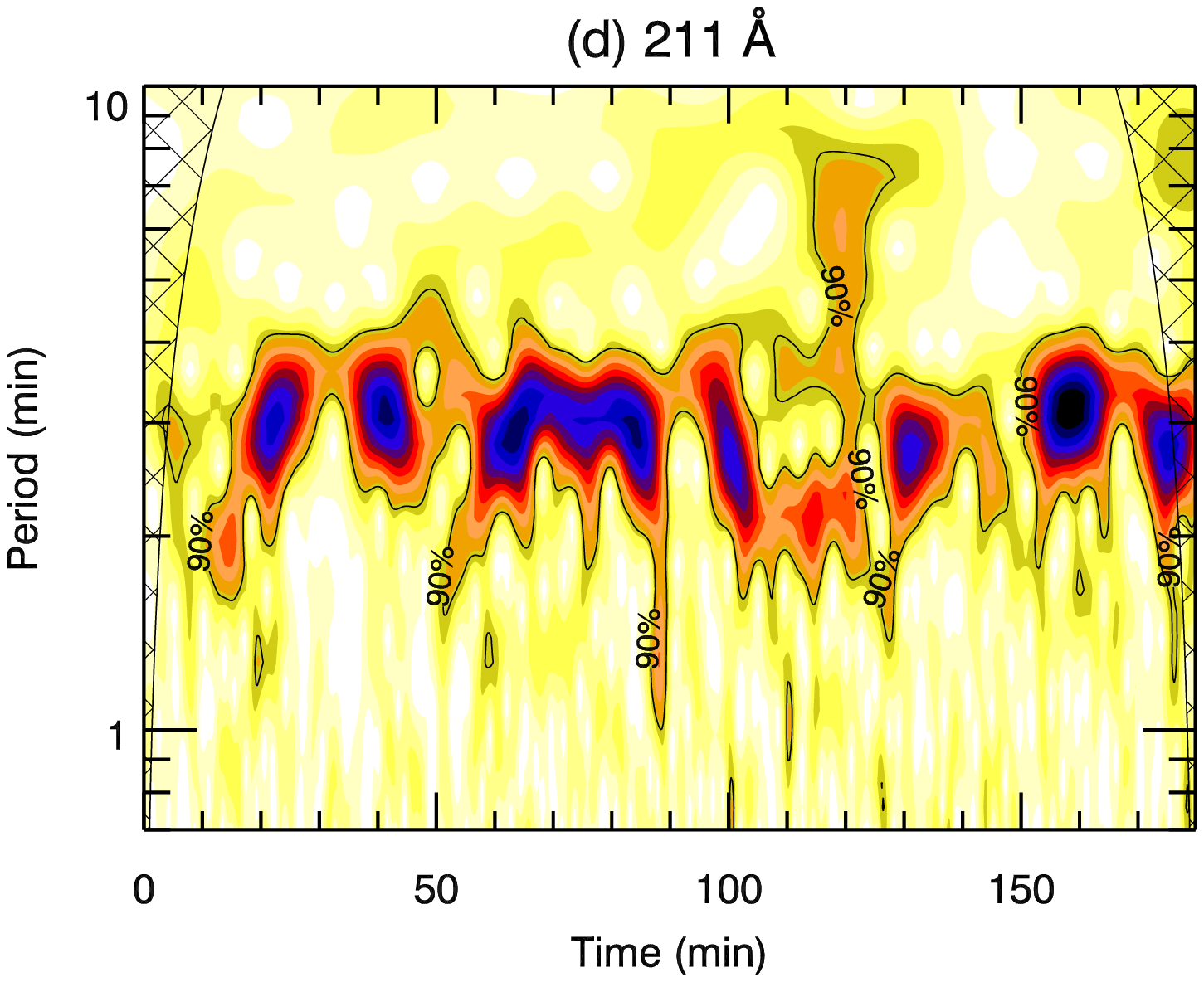}
\caption{Zoomed-up images of different wavelet power maps. Panel (a) is taken from right of Fig.~\ref{fig:waveletumbra}. Panel (b) is from left of Fig.~\ref{fig:wavelet171pix2_8}, and panels (c-d) are from Fig.~\ref{fig:wavelet193211}.} 
\label{fig:waveletharmonics}
\end{figure*}

 Besides the modulation, we observe some high-frequency oscillations around 1.5-min ($\approx$ 11 mHz) in the wavelet power maps (panel b). We show zoomed up versions of the wavelet power panels (b) in Fig.~\ref{fig:waveletharmonics} for different AIA channels as labelled. Such powers are observed at the footpoint of the fan loop in the sunspot umbra (Fig.~\ref{fig:umbraloc}) at 304~{\AA}, and at location L1 of the fan loops for the coronal channels 171~{\AA}, 193~{\AA}, and 211~{\AA}. However, they are not present throughout the time-series. Besides, we note that we detect the high-frequency components in 90\% confidence level. Therefore further observational work is required to establish the viability of this finding.
\section{Summary and Conclusions}\label{summary}
In this work, we have studied the nature of PIDs observed in an active region fan loop system rooted in a sunspot umbra using AIA/SDO observations. We find that the PIDs are best observed in 171~{\AA} data. The PIDs were also seen in other channels of AIA except for 335~{\AA} and 94~{\AA}. We have measured the phase speeds of the PIDs in all the AIA channels and found that the phase speeds are subsonic. This is suggestive that these are slow mode magneto-acoustic waves. We further found that the phase speeds measured in different channels are not dependent on temperature. This is in contradiction with the results obtained by, e.g., \citet{banerjee2009, KrishnaPrasad2012, 2012SoPh..279..427K, 2014A&A...568A..96G}. A possible reason for this may be the broad temperature response of the AIA channels. Fan loops are structures of typically 1 MK; therefore it may be that the emissions from the lower-temperature components are contributing to these different AIA EUV emissions \citep{2017ApJ...835..244G}, thereby giving similar phase speeds.

Our result shows that the PIDs have longer detection length ($\approx 5.9-7.35$ Mm) in 171~{\AA} and 193~{\AA}, whereas a shorter detection length ($\approx 2.9$ Mm) in 131~{\AA} and 211~{\AA}. The similar detection length in 171 and 193~{\AA} could be due to the similar temperature range covered by both these channels. We note that AIA 131~{\AA} has a peak formation temperature closer to that of 171~{\AA} in quiet conditions. Therefore, the shorter detection length observed in 131~{\AA} may be due to the weaker signal-to-noise ratio (SNR) (< 10) than that of 171~{\AA} (> 65). The shorter detection length observed in 211~{\AA} channel may also be related to its SNR, which we find to be less than 20.

The light curves obtained at multiple locations along the loops show an increase and decrease in the amplitude of oscillations over time. Such variation is better seen in the de-trended light curves (see Fig.~\ref{fig:or_dtred_lc}). Fouri\'er analysis on the original light curves reveals that there are several nearby frequencies within 5{--}8 mHz. The findings obtained here suggest that beat may be a possibility behind the observed amplitude modulation that are very likely created due to the interference of the nearby frequency components. We observe similar frequency peaks in all the AIA EUV coronal (except AIA 335~{\AA} and 94~{\AA}) as well as chromospheric channels. \citet{1986ApJ...301..992L} has reported the various frequencies peaks identified in our study in chromospheric umbra, and \citet{2001A&A...368..639F} reported them in transition region sunspot plumes. \citet{centeno2006, centeno2009, 2015ApJ...812L..15K} observed the presence of amplitude modulations in Four\'ier filtered light curves and measured a modulation period of 20-27 minutes \citep{2015ApJ...812L..15K}.
However, in our study, the amplitude modulations were prominent enough to detect them prior to any filtration (Fig.~\ref{fig:or_dtred_lc}), which to our knowledge is the first such observational finding. Wavelet analysis also reveals an increase and decrease of the oscillation power around 3-min simultaneous to the observed amplitude modulation. We obtain the modulation period to be in the range 20-30 minutes.

Observation of similar amplitude variation at the lower solar atmosphere (right panels of Fig.~\ref{fig:waveletumbra}) as observed by AIA 304~{\AA} (log [\textit{T/K}] = 4.7, \citet{2012SoPh..275...17L}) supports one more possibility that may be the amplitude variability is caused by the underlying driver. Photospheric \textit{p}-modes are, in general, assumed to be the cause of propagating slow magnetoacoustic waves observed in the corona \citep{2005ApJ...624L..61D, 2015ApJ...812L..15K, 2016ApJ...830L..17Z}. Hence, the amplitude variability of the \textit{p}-mode may affect and give rise to the particular amplitude variation of the propagating intensity waves observed along the fan loop. Observational studies have shown the effect oscillatory phenomenon in the lower solar atmosphere on the propagation of slow-magnetoacoustic waves along coronal fanloops \citep{2012ApJ...757..160J, 2017ApJ...850..206S}. However, further study is required both observationally and theoretically to understand the origin of such amplitude modulation in time series.

In addition to the amplitude modulation, we see presence of low period (high frequency) oscillations around 1.5-min (11 mHz) (see Fig.~\ref{fig:waveletharmonics}) in this study. We observe these high-frequency oscillations both at the chromospheric and coronal heights. \citet{Wang_2018} first observed high-frequency oscillations by imaging observations at different heights above sunspot umbra in the range 10 -- 14 mHz (around 1-min.). They related such high-frequency modes at coronal heights to perturbations at photospheric umbra. \citet{krishna2017ApJ} reported frequencies around 13 mHz using IRIS 2796~{\AA} in chromospheric sunspot umbra. We note that the observed high-frequency around 11 mHz is close to the double of the dominant (fundamental) frequency at around 5.56 mHz, and hence is similar to the second harmonics. However, one needs to be cautious to associate the high-frequency modes observed in our study to second harmonics, as harmonics are so far, observed in standing waves in coronal closed loops \citep{2010NewA...15....8S}. \citet{verwichteetal2004} reported the detection of higher harmonics in coronal kink wave oscillations with a deviation in the frequency ratio of second to first harmonics from its canonical value of 2 in a homogeneous medium. \citet{Andries_2005, Erdelyi_Verth2007, LunaCardozo2012ApJ} proposed that such deviation may occur because of longitudinal density and or magnetic stratification along stratified expanding flux tubes like those in the corona, and therefore can be used as a seismological tool to measure the density scale height.
\begin{acknowledgements} 
The authors are thankful to the referee for relevant suggestions on improving the manuscript. A.S. and D.T. acknowledge the Max-Planck Partner Group of the Max-Planck Institute for Solar System Research, G\"ottingen at IUCAA.
R.E. is grateful to Science and Technology Facilities Council (STFC grant nr ST/M000826/1) UK and to Chinese Academy of Sciences President’s International Fellowship Initiative, Grant No. 2019VMA0052  
for the support received. G.A. Ahmed acknowledges the visiting associateship of IUCAA. AIA and HMI data used here, are courtesy of SDO (NASA).
\end{acknowledgements}
\bibliographystyle{aa}
\bibliography{references}
\newpage
\begin{appendix}
\section*{Appendix A : True phase speeds of PIDs in AIA 171~{\AA}}
\begin{table}[!hbtp]
\centering
\caption{\label{tab3} True phase speeds of PIDs in AIA 171~{\AA} obtained by cosidering different inclination angle.}
\resizebox{7.5cm}{!} {
\begin{tabular}{lcc}
\hline
\hline
          &                    &  		 \\
{Projected Phase speed}     &  {Inclination angle}                         &	{Probable true }  \\   
in AIA 171~{\AA}            &                                              &Phase speeds \\
	     &                                              &                                   \\   
 {v [km~s$^{-1}$]}                           &	$\theta$ [$\degree$]                        &$v_{ph}$ [km~s$^{-1}$]	  \\    
\hline
\hline \space

           &     10              & 304.6           \\
           &    20              & 154.66            \\
52.9             &    30               & 105.8            \\ 
            &    40              & 82.29           \\ 
            &   50              & 69.05          \\ 
            &   60              & 61.08           \\ 
            &   70              & 56.29          \\ 
             &   80              & 53.71           \\ 
           &   90              & 52.9          \\ 
\hline
\hline
\end{tabular}
}
\end{table}

\section*{Appendix B : PIDs along slits 2 and 3}\label{slit2}
\begin{figure}[hbtp]
\centering
\includegraphics[trim=2.0cm 0.0cm 0.0cm 0.0cm,width=9.cm]{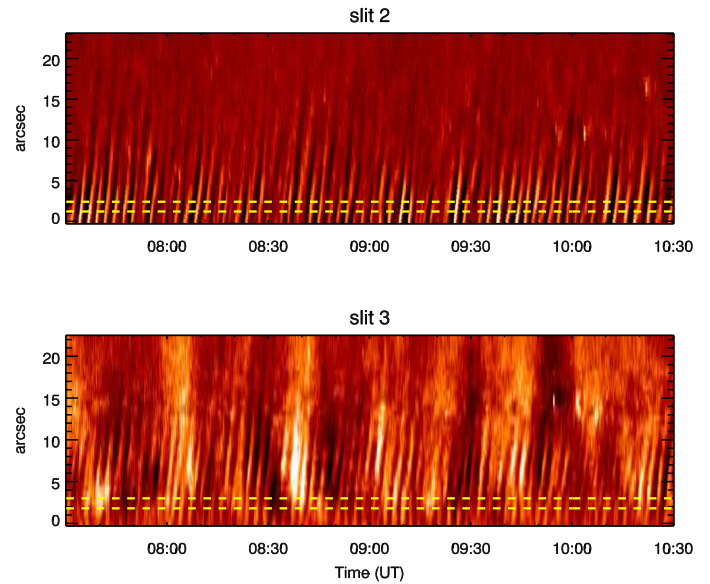}
\caption{Time-distance maps obtained along slits 2, and 3 in AIA 171~{\AA}. Clear propagation
of intensity oscillations along the slits are observed. The region averaged between
the overlaid yellow dashed lines are taken for further analysis.}
\label{fig:xts23}
\end{figure}
\begin{figure}[hbtp]
\centering 
\includegraphics[width=5.5cm,angle=90]{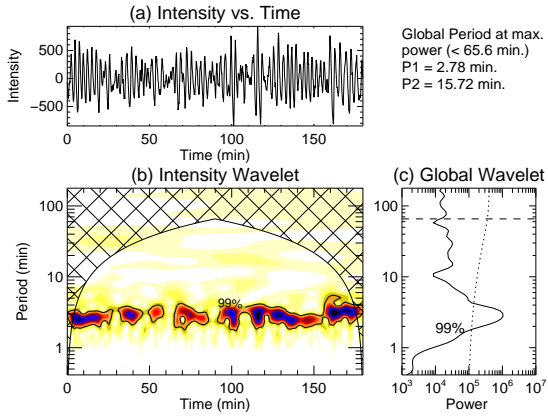}
\caption{Wavelet analysis of intensity disturbances propagating along slit 2. The light curve used for the wavelet analysis is obtained from the region averaged between the yellow dashed lines of Fig.~\ref{fig:xts23} (top panel:slit 2) in AIA 171~{\AA}. The different panels are explained in Fig.~\ref{fig:wavelet171pix2_8}.} 
\label{fig:wavelet_slit2}
\end{figure}
\begin{figure}[hbtp]
\centering 
\includegraphics[width=5.5cm,angle=90]{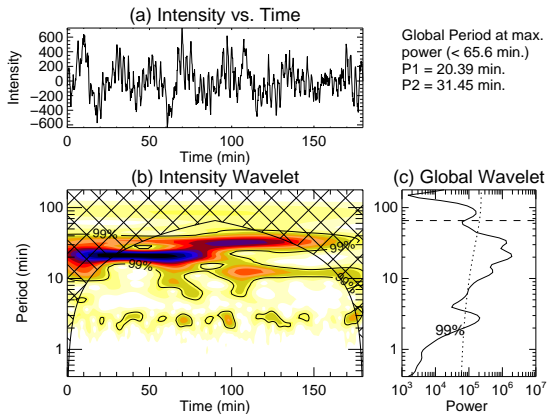}
\caption{Wavelet analysis of PIDs observed along slit 3 in AIA 171~{\AA}. The light curves shown in the top panel
correspond to the region averaged between the yellow dashed lines of Fig.~\ref{fig:xts23} (bottom panel:slit 3). The different panels are explained in  Fig.~\ref{fig:wavelet171pix2_8}.} 
\label{fig:wavelet_slit3}
\end{figure}
 We analyze the evolution of intensity disturbances propagating along slits 2 and 3 (shown in Figure~\ref{fig:lc}-a). We find the 3-min propagating intensity oscillations along slits 2 and 3 
 undergoing amplitude modulations. However, modulations were not strong enough as observed for slit 1. Figures~\ref{fig:wavelet_slit2} and \ref{fig:wavelet_slit3} present the increase and decrease of wavelet
power of the 3-min oscillations representing the modulations.

\end{appendix}

\end{document}